\documentclass[twocolumn]{aastex631}

\usepackage{booktabs}

\submitjournal{APJ}

\shorttitle{Theoretical approach to P chemistry in ISM}
\shortauthors{Fern\'andez-Ruz et al.}

\begin{document}

\title{A theoretical approach to the complex chemical evolution of phosphorus in the interstellar medium}

\author{Marina Fern\'andez-Ruz}
\affiliation{Centro de Astrobiolog\'{\i}a (CAB), CSIC-INTA, Ctra. de Ajalvir km 4, 28850 Torrej\'on de Ardoz, Spain}
\affiliation{Grupo Interdisciplinar de Sistemas Complejos (GISC), Madrid, Spain}

\author[0000-0003-4493-8714]{Izaskun Jim\'enez-Serra}
\affiliation{Centro de Astrobiolog\'{\i}a (CAB), CSIC-INTA, Ctra. de Ajalvir km 4, 28850 Torrej\'on de Ardoz, Spain}

\author[0000-0003-2196-5103]{Jacobo Aguirre}
\affiliation{Centro de Astrobiolog\'{\i}a (CAB), CSIC-INTA, Ctra. de Ajalvir km 4, 28850 Torrej\'on de Ardoz, Spain}
\affiliation{Grupo Interdisciplinar de Sistemas Complejos (GISC), Madrid, Spain}
\correspondingauthor{Jacobo Aguirre}
\email{jaguirre@cab.inta-csic.es}




\begin{abstract} 
The study of phosphorus chemistry in the interstellar medium has become a topic of growing interest in astrobiology, because it is plausible that a wide range of P-bearing molecules were introduced in the early Earth by the impact of asteroids and comets on its surface, enriching prebiotic chemistry. Thanks to extensive searches in recent years, it has become clear that P mainly appears in the form of PO and PN in molecular clouds and star-forming regions. Interestingly, PO is systematically more abundant than PN by factors typically of $\sim1.4-3$, independently of the physical properties of the observed source. In order to unveil the formation routes of PO and PN, in this work we introduce a mathematical model for the time evolution of the chemistry of P in an interstellar molecular cloud and analyze its associated chemical network as a complex dynamical system. By making reasonable assumptions, we reduce the network to obtain explicit mathematical expressions that describe the abundance evolution of P-bearing species and study the dependences of the abundance of PO and PN on the system's kinetic parameters with much faster computation times than available numerical methods. 
As a result, our model reveals that the formation of PO and PN is governed by just a few critical reactions, and fully explains the relationship between PO and PN abundances throughout the evolution of molecular clouds. Finally, the application of Bayesian methods constrains the real values of the most influential reaction rate coefficients making use of available observational data.

\end{abstract}

\keywords{Interstellar medium (847), Interstellar molecules (849), Astrochemistry (75), Astrobiology (74), Chemical reaction network models (2237), Bayesian statistics (1900)}

\section{Introduction} \label{sec:intro}

Phosphorus (P) is an essential element for life, being the fifth most abundant element in unicellular organisms, and the sixth in multicellular organisms \citep{MaciaBarber_2020}. P is present in phosphate groups, which can be found in several biomolecules including the informational polymers ribonucleic acid (RNA) and deoxyribonucleic acid (DNA), the phospholipids of the cell membrane, and the energetic molecules adenosine triphosphate (ATP) and guanosine triphosphate (GTP). Therefore, P must have played a key role in the early Earth prebiotic chemistry that, around 4 billion years ago, led to the origin of life in our planet. 

During the past decade it has been proposed that a significant part of the P reservoir on the early Earth surface might be of extraterrestrial origin \citep{Lefloch_2016,Rivilla_2016,Rivilla_2020,Bergner20,Bergner22}. Indeed, key volatile species such as PO have recently been detected in the comet 67P/Churyumov–Gerasimenko \citep{Altwegg16,Rivilla_2020}, supporting the hypothesis that comets and asteroids that fell abundantly onto our planet during the Late Heavy Bombardment period, enriched prebiotic chemistry with essential ingredients for the formation of the precursors of the building blocks of life, including P. In consequence, the astrobiological importance of P chemistry in the interstellar medium (ISM) relies on the fact that the chemical richness and complexity present in those comets was inherited from the chemistry occurred in the parental molecular cloud where our solar system was formed \citep{Altwegg16,Rivilla_2020,Bergner22}.

Interestingly, P is more scarce  at cosmic scales than other essential elements for life, such as H, C, O and N, something that has been named as `the phosphorus enigma' \citep{MaciaBarber_2020}. In fact, 
the number and complexity of the P-bearing molecules detected in space (in both the interstellar and the circumstellar medium) are still very limited: PO \citep{Tenenbaum_2007,Rivilla_2016}, PN \citep{Ziurys_1987,Fontani16}, CP \citep{Guelin_1990}, HCP \citep{Agundez_2007}, CCP \citep{Halfen_2008}, PH$_3$ \citep{Agundez08,Agundez14a}, NCCP \citep{Agundez_2014b}, and PO$^+$ \citep{Rivilla_2022}.

In recent years, PO and PN have attracted special attention among the astrochemistry community because they are the only P-bearing species that have been detected in
molecular clouds and star forming regions \citep[see e.g.][]{Ziurys_1987,Fontani16, Rivilla_2016, Lefloch_2016, Rivilla_2018, Rivilla_2020, Bernal_2021, Bergner19, Bergner22}. All these observational works reveal that PO is systematically more abundant than PN with abundance ratios of [PO]/[PN] $\sim1.4-3$, independently of the observed source. These ratios are, however, not easy to reproduce by existing astrochemical models since they predict [PO]/[PN] ratios $<1$ for a wide range of physical conditions \citep{Jimenez-Serra_2018,Chantzos_2020,Sil_2021}. This implies that, despite the abundant information available from astronomical observations, the formation routes of PO and PN remain unclear. The inconsistency between the models and the observations may be due to several reasons: (i) the incompleteness of the chemical network of P; (ii) the large uncertainties in the reaction rate coefficients that models have to deal with; and (iii) the unknown yields of surface reactions on grains, which determine the main reservoir of solid P and the form in which P is made available in the gas phase.

For point (i), \citet{Jimenez-Serra_2018} proposed that the reaction P+OH $\rightarrow$ PO+H, missing in astrochemical models, could be an efficient mechanism of formation of PO. This has been recently confirmed by \citet{GarciadelaConcepcion_2021}, who have performed quantum-chemical and kinetic calculations of this reaction proving that it is indeed one of the main formation routes of PO in magnetohydrodynamic shocks with shock speeds $\geq$40 km s$^{-1}$. 

For points (ii) and (iii), astrochemical codes have to deal with the already mentioned uncertainty of the reaction rate coefficients needed to solve the system of ordinary differential equations (ODEs) associated with a big set of reactions. The majority of the reaction rates are tabulated in databases such as UMIST \citep{McElroy13} and KIDA \citep[KInetic Database for Astrochemistry;][]{Wakelam_2012}, but many of these values have not been validated with theoretical or experimental methods. In addition, little is known about the main reservoir of P on dust grains, although it is suspected that it resides in a semi-refractory form \citep{Bergner22}. 

In this work, we introduce a mathematical model of the evolution of P chemistry in the ISM to cast light on the formation routes of PO and PN. We analyze the dependences of the PO and PN abundances on the reaction rate coefficients and the influence of the main reservoir of P on grains. By making  appropriate assumptions, we reduce the number of reactions of the complex chemical network of phosphorus to the key reactions that are then analyzed. The simplicity of our approach allows us to obtain explicit mathematical expressions that reproduce the abundance evolution with time of the P-bearing species involved in our reduced P-network. Such complete analytical solution of the system permits much faster computation times than currently available numerical methods from astrochemical codes. Taking advantage of this method, we analyze in detail the dependence of the model on the parameter space, (i) showing the main pathways by which PO and PN are formed and/or destroyed to provide a general explanation to why [PO]/[PN] values are systematically $<$1 in models but $>$1 in real data; and (ii) identifying the most critical reaction rate coefficients involved in the process. Finally, the application of Bayesian statistics is used in the refinement of the calculation of such coefficients with the aid of observational data available in the literature.

\begin{deluxetable*}{ccccccccc}\label{tab:set_of_reactions}
\tablecaption{Set of chemical reactions, kinetic parameters $\alpha$, $\beta$ and $\gamma$ of the modified Arrhenius equation, reaction rate coefficients $k_j$ (for temperatures 10 K, 100 K and 300 K) and bibliographical sources.}
\tablehead{\colhead{$j$} & \colhead{Reaction} & \colhead{$\alpha$} & \colhead{$\beta$} & \colhead{$\gamma$} &\colhead{$k_{j}$ ($T$=10 K)} & \colhead{$k_{j}$ ($T$=100 K)} & \colhead{$k_{j}$ ($T$=300 K)} & \colhead{References}  \\
\colhead{} & \colhead{} & \colhead{(cm$^{3}$ s$^{-1}$)} & \colhead{} & \colhead{} & \colhead{(cm$^{3}$ s$^{-1}$)} & \colhead{(cm$^{3}$ s$^{-1}$)} & \colhead{(cm$^{3}$ s$^{-1}$)} & \colhead{} } 
\startdata
1 & N+PO $\rightarrow$ P+NO & $2.55\times 10^{-12}$ & 0 & 0 & $2.55\times 10^{-12}$ & $2.55\times 10^{-12}$ & $2.55\times 10^{-12}$ & 1 \\
2 & N+PO $\rightarrow$ PN+O & $3.00\times 10^{-11} $ & -0.6 & 0 & $2.31\times 10^{-10} $ & $5.80\times 10^{-11} $ & $3.00\times 10^{-11} $ & 1 \\
3 & O+PH$_2$ $\rightarrow$ PO+H$_2$ & $4.00\times 10^{-11} $ & 0 & 0 & $4.00\times 10^{-11} $ & $4.00\times 10^{-11} $ & $4.00\times 10^{-11} $ & 1\\
4 & O+PH $\rightarrow$ PO+H  & $1.00\times 10^{-10} $ & 0  & 0 & $1.00\times 10^{-10} $ & $1.00\times 10^{-10} $ & $1.00\times 10^{-10} $ & 1 \\
5 & P+O$_2$ $\rightarrow$ PO+O & $3.99\times 10^{-12} $ & 0.89  & 814 & $8.61\times 10^{-49} $  &  $4.38\times 10^{-16} $ &  $2.65\times 10^{-13} $ & 4 \\
6 & P+OH $\rightarrow$ PO+H  & $2.28\times 10^{-10} $ & 0.16 & 0.37 & $1.28\times 10^{-10} $ & $1.91\times 10^{-10} $ & $2.28\times 10^{-10} $ & 2 \\
7 & N+PH $\rightarrow$ PN+H  & $8.80\times 10^{-11} $& -0.18 & 1.01 & $1.47\times 10^{-10} $ & $1.06\times 10^{-10} $ & $8.77\times 10^{-11} $ & 5 \\
8 & N+CP $\rightarrow$ PN+C  & $8.80\times 10^{-11} $ & 0.42 & 0 & $2.11\times 10^{-11} $ & $5.55\times 10^{-11} $ & $8.80\times 10^{-11} $ & 1$^{a}$ \\
9 & P+CN $\rightarrow$ PN+C     & $8.80\times 10^{-11} $ & 0.42 & 0 & $2.11\times 10^{-11} $ & $5.55\times 10^{-11} $ & $8.80\times 10^{-11} $ & 1$^{a}$ \\
10 & H+PH $\rightarrow$ P+H$_2$  & $1.50\times 10^{-10} $ & 0 & 416 & $1.29\times 10^{-28} $ & $2.34\times 10^{-12} $ & $3.75\times 10^{-11} $ & 3 \\
11 & O+CP $\rightarrow$ P+CO   & $4.00\times 10^{-11} $ & 0 & 0 & $4.00\times 10^{-11} $ & $4.00\times 10^{-11} $ & $4.00\times 10^{-11} $ & 1 \\
12 & H+PH$_2$ $\rightarrow$ PH+H$_2$  & $6.20\times 10^{-11} $ & 0 & 318 & $9.59\times 10^{-25} $ & $2.58\times 10^{-12} $ & $2.15\times 10^{-11} $  & 3 \\
13 & H+PH$_3$ $\rightarrow$ PH$_2$+H$_2$  & $4.50\times 10^{-11} $ & 0 & 735 &  $5.40\times 10^{-43} $ & $2.89\times 10^{-14} $ & $3.88\times 10^{-12} $ & 3 \\
14 & C+PH $\rightarrow$ CP+H & $7.50\times 10^{-11} $ & 0 & 0 & $7.50\times 10^{-11} $ & $7.50\times 10^{-11} $ &  $7.50\times 10^{-11}$ & 1 \\
\enddata
\tablecomments{$^{a}$ No bibliography was available for those reactions, so we used the parameters associated with the analogous Nitrogen (N) reaction N+CN $\rightarrow$ N$_2$+C, as previous works suggest that the chemical similarity between N and P could lead to a similar chemical behavior \citep{Agundez_2007}.}
\tablerefs{(1) KIDA \cite{Wakelam_2012}, (2) \cite{GarciadelaConcepcion_2021}, (3) \cite{Charnley_1994}, (4) \cite{GarciadelaConcepcion_2023} (5) \cite{Gomes_2023}.}
\end{deluxetable*}
\begin{figure*}
\centering
\includegraphics[width=\textwidth]{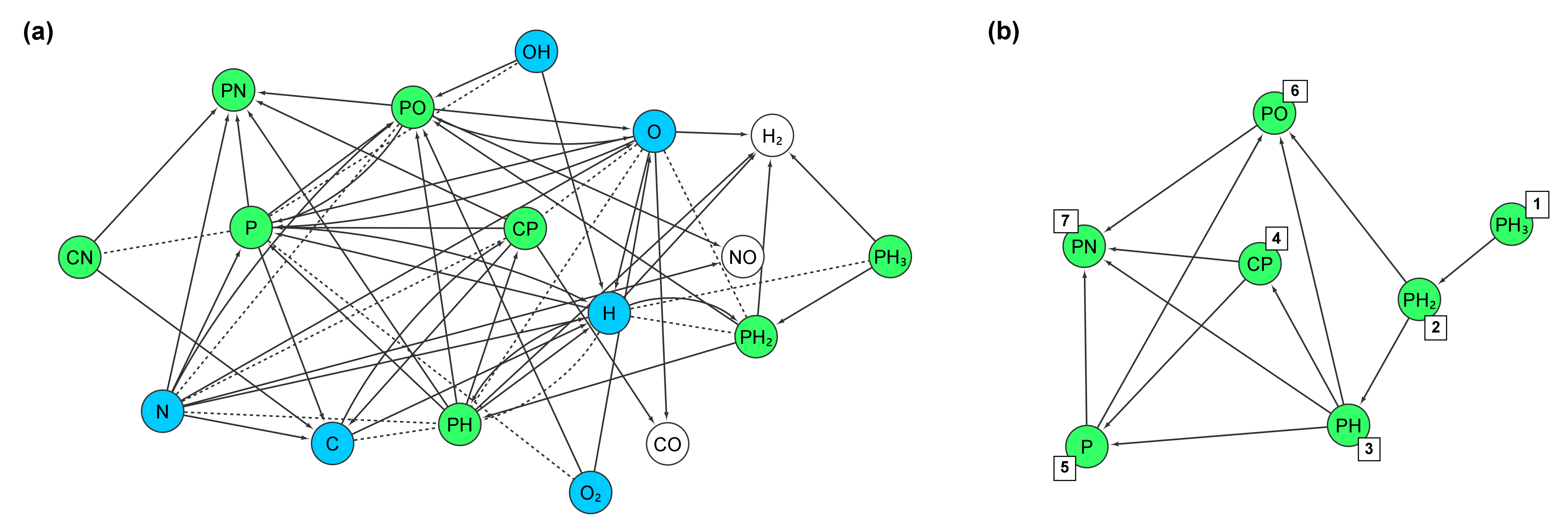}
\caption{Complex networks associated with the set of reactions selected from the chemistry of phosphorus in the ISM for our model. (a) Chemical network representing the 17 species and 14 chemical reactions analyzed in this work. Nodes represent chemical species involved in the system, and are classified as follows: abundant (blue nodes), scarce (green) and non-interacting (white) according to the criterion explained in the text. Directed links (arrows) go from the reactants to the products of a reaction and undirected links (dashed lines) connect the reactants of a reaction. (b) Sub-network of the total network plotted in (a) that sketches the theoretically solved system. We have included the indices $i$ used in Equations~(\ref{eq:mass_action_theory}-\ref{eq:general_solution}) to number the P-bearing species in the minimal system. \label{fig:networks}} 
\end{figure*}
\section{A model for the chemical evolution of phosphorus in the interstellar medium}\label{theoreticalframework}

The theoretical model involves 17 chemical species and is composed of a reduced set of 14 chemical reactions (see Table~\ref{tab:set_of_reactions}) that are assumed to take place in the ISM. To generate this set of chemical reactions, we have taken the chemical network built by \citet{Jimenez-Serra_2018} and recently augmented by \citet{GarciadelaConcepcion_2021}. This chemical network considers all reactions with P-bearing species present in the UMIST database \citep{McElroy13}, which includes the original network for phosphorus of \cite{Millar_1991}, plus additional reactions extracted from \cite{Charnley_1994}, \cite{Anicich_1993} and \cite{Agundez_2007} \citep[see][for details]{Jimenez-Serra_2018}. The chemical network also includes the newly calculated rate constants for the reactions P+OH $\rightarrow$ PO+H \citep{GarciadelaConcepcion_2021}, P+O$_2$ $\rightarrow$ PO+O (Garc\'{\i}a de la Concepci\'on et al. 2023, submitted) and N+PH $\rightarrow$ PN+H \citep{Gomes_2023}.
In this work, we focus on neutral-neutral gas-phase reactions because, unlike ion-neutral reactions \citep[see e.g.][]{Thorne_1984}, the neutral-neutral ones have not been measured in the laboratory and therefore they are subject to large uncertainties \citep[most reaction rates are best guesses due to the difficulties in performing laboratory experiments with these species; see e.g.][]{Millar_1987}. In addition, these reactions are expected to dominate the chemistry of P-bearing molecules in the regions where PN and PO have been found since the ionization fraction of the gas is low \citep{Jimenez-Serra_2018}. 
Indeed, ion-neutral and dissociative recombination reactions are known to be minor contributors to the formation of PO and PN in molecular clouds and star-forming regions from previous theoretical studies \citep{Millar_1987,Charnley_1994}, unless an extremely high UV radiation field or cosmic rays ionization rate is present \citep{Jimenez-Serra_2018,Rivilla_2022}. However, here we only focus on deeply embedded star-forming regions, where most detections of PO and PN have been reported. In regions where photochemistry is relevant, the chemical network could not be reduced as we do it here. 

The selected set of reactions is finally represented as a complex network (see Figure~\ref{fig:networks}(a)), where nodes are the 17 chemical species involved, directed links (arrows) go from the reactants to the products of the same reaction, and undirected links (dashed lines) connect both reactants of a reaction.

The chemistry is modeled according to the law of mass action \citep{Chang_2017}. Consequently, given a set of reactions of the form A+B $\rightarrow$ C+D, the rate of change with time of the abundance of each chemical species $i$ is given by
\begin{equation}
\centering
\frac{d[X_i]}{dt}=\sum_{l,m} k_{lm}^{i} n_{\mathrm{H}} [X_l] [X_m] - [X_i] \sum_n   k_{ni}n_{\mathrm{H}} [X_n]\,,
\label{eq:mass_action}
\end{equation}
where $[X_i]$ is the abundance of species $i$ relative to the abundance of H, and $n_{\mathrm{H}}$ is the H number density. The first sum contains the formation terms and the second sum contains the destruction terms of species $i$. $k_{lm}^{i}$ are the reaction rate coefficients of the reactions between the reactants $X_l$ and $X_m$ that produce species $i$ (i.e. $X_l+X_m\rightarrow X_i+X_n$), while $k_{ni}$ is the reaction rate coefficient of all the reactions in which species $X_i$ is a reactant (i.e. $X_n+X_i\rightarrow \mathrm {products}$). If we apply Equation~(\ref{eq:mass_action}) to the 17 chemical species in the network, we obtain the associated system of ODEs explicitly shown in Appendix~\ref{appendix:ODEs} whose solution describes the evolution with time of the abundances of all molecules.  
\begin{deluxetable}{cccc}\label{tab:initial_abundances}
\tablecaption{Initial abundances with respect to H of the species involved in the phosphorus chemistry network in the ISM studied in this work.}
\tablehead{\colhead{Species} & \colhead{Initial abundance} & \colhead{Type} & \colhead{References}} 
\startdata
C     & $2.69 \times 10^{-4}$                       & A    &      1  \\
CN    & $5.92 \times 10^{-10}$                 & S    &      2$^{a}$ \\
CO    & -                                      & NI $^{e}$   &      N/A  \\
CP      & $1.00 \times 10^{-13}$                 & S    &     N/A$^{b}$ \\
H      & 1                                      & A    &       N/A$^{c}$ \\
H$_2$   & -                                      & NI $^{e}$   &      N/A  \\
N       & $6.76 \times 10^{-5}$                  & A    &       1 \\
NO   & -                                      & NI $^{e}$   &       N/A \\
O       & $4.90 \times 10^{-4}$                  & A    &      1  \\
O$_2$   & $6.04 \times 10^{-7}$                  & A    &      3,4 $^{a}$\\
OH      & $1.00 \times 10^{-7}$                   & A    &     5  \\
P & $(1-f_\mathrm{P})\times 2.57 \times 10^{-9} $ & S$^{d}$ & 1 \\
PH  & $(f_\mathrm{P}/ 3) \times 2.57 \times 10^{-9}$ & S$^{d}$ & 1\\
PH$_2$ & $(f_\mathrm{P}/ 3) \times 2.57 \times 10^{-9}$ & S$^{d}$ & 1\\
PH$_3$ & $(f_\mathrm{P}/ 3) \times 2.57 \times 10^{-9}$ & S$^{d}$ & 1\\
PN      & 0                                      & S    &     N/A $^{c}$  \\
PO      & 0                                      & S    &     N/A $^{c}$   \\
\enddata
\tablecomments{$^{a}$ In cases where the source provided two values or we considered two sources, we used the geometric mean.\\$^{b}$ Up to date, CP has not been detected in the ISM, but it has been detected in a circumstellar shell envelope by \cite{Guelin_1990}. Thus, in our model we consider that CP is present but we fix its initial value to $10^{-13}$ so it is  sufficiently below the detection limit ($\sim 10^{-12}) $.\\$^{c}$ The value is set to one (for H) and zero (for PO and PN) following the model's rules.\\$^{d}$ The initial abundances of atomic P, PH, PH$_2$ and PH$_3$ with respect to H are expressed in terms of the P-hydrogenation fraction $f_\mathrm{P}$.\\$^{e}$ The initial abundances of non-interacting (NI) species are not needed to solve numerically or theoretically the rest of the system.}
\tablerefs{(1) \cite{Jimenez-Serra_2018}, (2) \cite{Agundez_2013}, (3) \cite{Larsson_2007}, (4) \cite{Goldsmith_2011}, (5) \cite{Rugel_2018} } \ 
\end{deluxetable}

For neutral-neutral gas-phase reactions, the most common form to parameterize the dependence of the reaction rate coefficient on temperature is given by the modified Arrhenius equation,

\begin{equation}\label{eq:arrhenius}
    k(T)=\alpha \left( \frac{T}{300} \right)^{\beta} \mathrm{exp}\left( -\frac{\gamma}{T} \right)\,,
\end{equation}
where $\alpha$, $\beta$ and $\gamma$ are the kinetic parameters and T is the temperature. The kinetic parameters $\alpha$, $\beta$ and $\gamma$ of the 14 reaction rate coefficients have been obtained from the sources specified in Table~\ref{tab:set_of_reactions} and have been used to obtain the rate coefficients $k_j$ of each reaction $j$.

Our model takes into account the initial abundances of the chemical species involved in the chemical network. We have assumed solar abundances for the atomic species \citep[extracted from][]{Asplund_2009,Jimenez-Serra_2018}, while for the molecules O$_2$, CN, and OH, we use the abundances measured toward molecular clouds (see Table~\ref{tab:initial_abundances} and references therein). For CP, since it has not been detected in the ISM, we just assume an abundance below the typical detection limit of $\sim$10$^{-12}$. All initial conditions can be found in Table~\ref{tab:initial_abundances} along with their bibliographic sources.

The chemical species are classified according to their initial abundance in one of the following groups: abundant (A), scarce (S) or non-interacting species (NI). Throughout the paper we assume that \textit{abundant} species are those whose initial abundance is greater than $10^{-8}$ with respect to H, and are represented as blue nodes in the chemical network plotted in Figure~\ref{fig:networks}. \textit{Scarce} (S) chemical species are those whose initial abundance is below $10^{-8}$, and are represented as green nodes. Finally, there are three species that are not reactants of any reaction and therefore their abundances do not appear in the right side of any equation in the system of ODEs introduced in Appendix~\ref{appendix:ODEs}. Independently of their initial abundance, they have been called \textit{non-interacting} (NI) species, and are represented as white nodes. 

While the model describes explicitly the gas-phase chemistry, the grain-surface chemistry is also implicitly included as follows. It has been argued that the observed scarcity of P in the gas phase in molecular clouds is because most of the atomic P freezes out onto dust grains \citep{Ziurys_1987,Turner87,Aota12,Lefloch_2016}. In accordance with this assumption, the sum of the initial abundances of atomic P, PH, PH$_2$ and PH$_3$ in the model has been depleted by a factor of 100 with respect to the cosmic abundance of P, proceeding as in previous works \citep[see e.g.][]{Aota12,Lefloch_2016,Jimenez-Serra_2018}. 
In addition, it is believed that molecules PH, PH$_2$ and PH$_3$ are formed on the dust grain
surfaces through hydrogenation of atomic P \citep{Charnley_1994} before being released to the gas phase, but the 
actual yields of the surface reactions transforming P into PH, PH$_2$ and PH$_3$ are unknown. In order to account for this uncertainty in our simulations, we define the P-hydrogenation fraction $f_{\mathrm{P}}$ as the fraction of P that is initially in the form of PH, PH$_2$ and PH$_3$. The initial abundances of P, PH, PH$_2$ and PH$_3$ depend on $f_{\mathrm{P}}$ as described in Table~\ref{tab:initial_abundances}. 
As an example, in our model $f_{\mathrm{P}}=0$ means that all initial P is in the form of atomic P, while $f_{\mathrm{P}}=1$ means that all initial P is hydrogenated and equally distributed between PH, PH$_2$ and PH$_3$. 

\section{Analysis of the system}
\subsection{Theoretical Solution}\label{subsec:theoretical_solution}

The set of ODEs that describes the chemical evolution of phosphorus in the interstellar medium can be mathematically solved under certain approximations that transform the system of 17 nonlinear ODEs into a minimal linear system of 7 ODEs corresponding to the P-bearing species P, PH, PH$_2$, PH$_3$, CP, PO and PN. 
Note that, for the sake of clarity, we will name throughout the paper {\it total system} to the one composed of the 17 nonlinear ODEs (Equations~(\ref{eq:total_system_begin}-\ref{eq:total_system_end}) in Appendix~\ref{appendix:ODEs}) and {\it minimal system} to the mathematically solvable system made of 7 linear ODEs (set of Equations~(\ref{Eq:reduced}) in Appendix~\ref{appendix:analytical}). 
To obtain the minimal system and be able to solve it mathematically, we must assume that (i) the abundant species are constant for all times (i.e. $d[X]/dt=0$ for $X=$ C, H, N, O, O$_2$, OH), (ii) CN abundance is constant as its rate of change, $d[\mathrm{CN}]/dt = -k_{9} n_{\mathrm{H}}[\mathrm{P}][\mathrm{CN}]$, is extremely small because both P and CN are scarce species, and (iii) the term $k_1[\mathrm{N}][\mathrm{PO}]$ in Equation~(\ref{Eq:P}) is negligible because its value is several orders of magnitude lower than the dominant terms and in consequence the same applies to the arrow from PO to P in Figure~\ref{fig:networks}(b) (see Appendices~\ref{appendix:analytical}  and \ref{appendix:approximation} for a thorough analysis of the suitability of these assumptions).

Furthermore, as the non-interacting species NO, H$_2$ and CO do not influence the evolution of the rest of molecules and we are only interested in the evolution of the P-bearing species, we can neglect their kinetic equations and finally obtain an independent set of seven ODEs for the minimal system, where every differential equation is linear and of the type 
\begin{equation}\label{eq:mass_action_theory}
    \frac{d[X_i]}{dt}=\sum_{j \neq i}k_j n_{\mathrm{H}} [Y_j] [X_j] - [X_i] \sum_{j} k_j n_{\mathrm{H}} [Y_j]\,,
\end{equation}
where $X$ stand for the P-bearing species, $Y$ for the non P-bearing species, and $i$ numbers the species according to Figure~\ref{fig:networks}(b). The right-hand first sum and second sum are the formation and the destruction terms of P-bearing species $i$, respectively. Note that non P-bearing species $Y_j$ belong to the abundant type (A) and therefore verify $[Y_j]=[Y_j]_0$ for all times, while P-bearing species $X_j$ belong to the scarce type and verify $[X_j]_0\ll[Y_j]_0$ (as mentioned above, CN is also scarce but was treated differently). 

While obtaining the explicit solution of a linear ODE system with seven equations is in general unfeasible, in this case we can do it by solving the equations sequentially, as the matrix of coefficients associated with the system is triangular. 
This property has a graphic counterpart in the fact that the sub-network (of the total chemical network) shown in Figure~\ref{fig:networks}(b) composed of the 7 P-bearing species and the links connecting them does not have any cycles, that is, if we start a walk in any of those nodes, there are no paths to go back to the original node by following the directed links of the network. 

Making the mentioned assumptions and following the steps described above, we obtain a general explicit expression for the time-evolution of the abundances $[X_i]$ of each P-bearing species $i$,
\begin{equation}\label{eq:general_solution}
    [X_i](t)=\left[ \sum^{i-1}_{j=1} \frac{C_{ij}}{r_i-r_j}\,e^{-r_j\,t}\right]+C_{ii}\,e^{-r_i\,t}\,,
\end{equation}
where $C$ and $r$ are constants that depend on the reaction rate coefficients and the initial abundances and whose expressions are given in Appendix~\ref{appendix:analytical}.  
Note that constant $r_i$ represents the decay velocity of the consumption of species $i$ due to its own interaction with other species. A clarifying example: CP ($i=4$) is consumed in reactions 8 and 11, interacting with N and O respectively (remarked in dashed lines in Figure~\ref{fig:networks}(a)). Its associated decay velocity is then $r_4=k_{8} n_{\mathrm{H}} [\mathrm{N}]_0 + k_{11} n_{\mathrm{H}} [\mathrm{O}]_0\,.$

Appendix~\ref{appendix:analytical} shows the complete mathematical derivation of the solutions of the minimal system introduced in Equation~(\ref{eq:general_solution}) and described above. Furthermore, in Appendix~\ref{appendix:approximation} we assess the rightness and caveats of assuming constant the abundance of the species classified as abundant, a necessary premise to obtain the theoretical solution. In particular, we provide a theoretical calculation where we show that, for chemical reactions of the type $\mathrm{A} + \mathrm{B} \rightarrow \mathrm{C} + \mathrm{D}$, the error in the calculation of the evolution of A and B when assuming that the abundance of B is constant for the times analyzed in this work ($t\leq 10^5$ yrs) becomes negligible when the initial conditions verify $B_0>>A_0$. Finally, in Appendix~\ref{appendix:ratio} we analyze theoretically the ratio [PO]/[PN] for the first stage of the chemical evolution of the system, in order to cast light on the [PO]/[PN] disagreement between models and observational data.

\subsection{Numerical Solution}\label{subsec:numerical}

\begin{figure*}
\centering
\includegraphics[width=\textwidth]{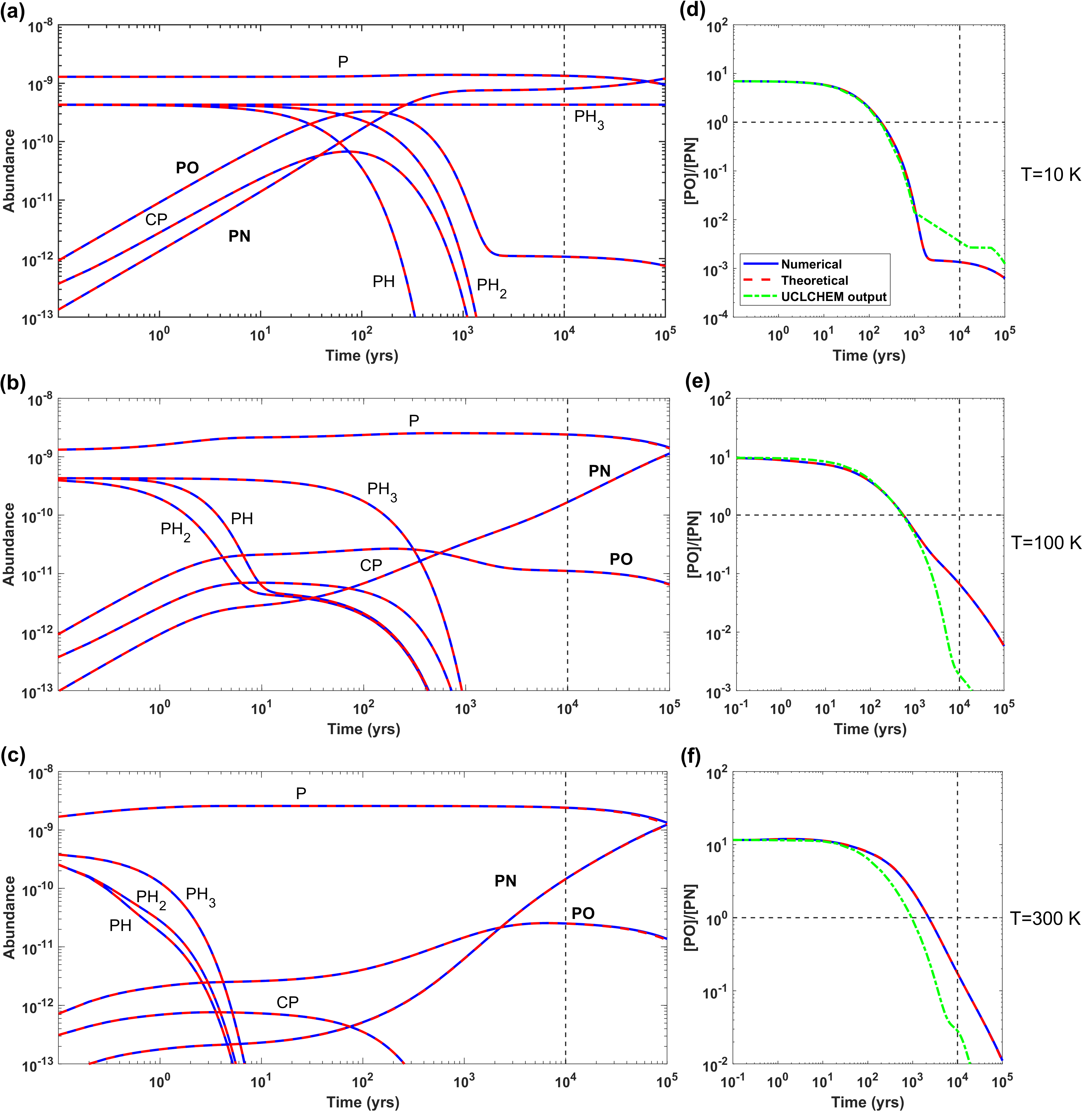}
\caption{Evolution of the abundances relative to H of the P-bearing molecules (a-c) and the ratio [PO]/[PN] (d-f) for $T$=10 K, 100 K, and 300 K respectively. P-hydrogenation fraction $f_\mathrm{P}=0.5$ (i.e. 50\% of initial P locked into atomic P and 50\% equally distributed between PH, PH$_2$ and PH$_3$) in all cases. Results were obtained solving the total model numerically (blue lines) and through the theoretical solution of the minimal system (dashed red lines). Note that the numerical and theoretical approaches yield identical results to the naked eye for all species, times and temperatures. A dashed vertical line remarks the typical cloud age, $t=10^4-10^5$ yrs, and a dashed horizontal line is marked at  [PO]/[PN]=1 in (d-f).
For comparison, the curve of [PO]/[PN] obtained from UCLCHEM chemical code \citep{Holdship_2017} has been plotted in (d-f) (dash-dotted green lines). \label{fig:numerical_evolution}}
\end{figure*}

Throughout this work, we will model three typical astrophysical scenarios where reactions can take place: a molecular cloud during the cold collapse phase (at $T$=10 K) and a star-forming region affected by shocks with average gas temperatures of $T$=100 K and $T$=300 K. For all of them, simulations are performed for time-scales of $10^5$ yrs \citep[see e.g.][]{Fontani16,Jimenez-Serra_2018} and assuming that the cloud density is constant with time, with the H number density $n_{\mathrm{H}}=10^4$ cm$^{-3}$. Note that our aim is not to reproduce the astrochemical modeling done in previous works \citep[where typically multiple evolutionary phases/stages are considered; see e.g.][]{Aota12, Lefloch_2016, Jimenez-Serra_2018}, but to analyze in detail the dependences of the [PO]/[PN]
abundance ratio on the assumed reaction rate coefficients, and to understand why this ratio is systematically $< 1$ in models but $>1$ in observational data.

In Figure~\ref{fig:numerical_evolution}(a-c) the evolution curves of P, PH, PH$_2$, PH$_3$, CP, PO and PN abundances are plotted for $T$=10 K, $T$=100 K, and $T$=300 K, with a P-hydrogenation fraction $f_\mathrm{P}=0.5$ (i.e. 50\% of the initial P locked into atomic P and 50\% equally distributed between PH, PH$_2$ and PH$_3$).  In Figure~\ref{fig:numerical_evolution}(d-f) the ratio [PO]/[PN] is represented under the same conditions.
The abundances have been calculated applying numerical methods (the total system with 17 equations --Appendix~\ref{appendix:ODEs}-- has been solved applying a fourth-order Runge-Kutta algorithm with a constant time step of 0.1 yr) and through the theoretically obtained expressions for the minimal system introduced in Equation~(\ref{eq:general_solution}). To ensure that the chemical system studied here is not significantly affected by the lack of ion-neutral and dissociative recombination reactions, in Figure~\ref{fig:numerical_evolution}(d-f) we also show the [PO]/[PN] ratio obtained with the astrochemical code UCLCHEM \citep{Holdship_2017} using the same physical conditions and initial abundances of Table$\,$\ref{tab:initial_abundances}. From Figure~\ref{fig:numerical_evolution}(d-f), it is clear that the [PO]/[PN] ratios derived using UCLCHEM are in perfect agreement with the ones derived using our model for time-scales$\leq$1000 yrs for $T$=10 and 100 K, and for time-scales$\leq$100 yrs for $T$=300 K. For time-scales larger than these, our model predictions deviate from the UCLCHEM's results. However, note that the evolutionary trends are preserved and thus, these discrepancies do not qualitatively affect our conclusions.

For $T$=10 K, PH and PH$_2$ are initially transformed into PO, PN and CP as a result of reactions 3 (O+PH$_2$ $\rightarrow$ PO+H$_2$), 4 (O+PH $\rightarrow$ PO+H), 7 (N+PH $\rightarrow$ PN+H) and 14 (C+PH $\rightarrow$ CP+H). Although the initial abundances of PO and PN are zero, and CP very scarce, after a few hundreds of years all three have reached detectable abundances of about 10$^{-11}$-10$^{-10}$ (note that fixing the initial CP abundance to zero would yield almost indistinguishable results). In a second stage of the evolution, PH and PH$_2$ get depleted (but not PH$_3$) at $t\sim10^3$ yrs and consequently reactions 3, 4, 7 and 14 become negligible, resulting in a strong decay of CP and PO. PN gets strongly reinforced from there on as PO has become abundant enough to enhance reaction 2 (N+PO $\rightarrow$ PN+O), a reaction that was negligible in the first stage of the evolution of the system. This transformation of PO into PN beyond $t\sim10^3$ yrs reinforces the decrease in the [PO]/[PN] ratio, which drops from its initial value $\sim 7$ to $\sim 0.02$ at around $t\sim10^3$ yrs.   
 Note that at this temperature, reaction 6 (P+OH $\rightarrow$ PO+H) is not strong enough to prevent PO from being consumed, but it strongly slows down its decrease and that of the ratio [PO]/[PN] for long times.

For higher temperatures such as $T$=100 K and $T$=300 K, the route PH$_3\rightarrow$ PH$_2\rightarrow$ PH $\rightarrow$ P is activated since the rate coefficients of chain reactions 13 (H+PH$_3$ $\rightarrow$ PH$_2$+H$_2$), 12 (H+PH$_2$ $\rightarrow$ PH+H$_2$) and 10 (H+PH $\rightarrow$ P+H$_2$) have a strong positive dependence on temperature (they are endothermic; see Table~\ref{tab:set_of_reactions}). This phenomenon results in a constant growth of P and a fast depletion of PH, PH$_2$ and PH$_3$ (at $T$=100 K these species get depleted in the first $10^3$ yrs of evolution, and at $T$=300 K the process is even faster). The growth of P reinforces reaction 6 (P+OH $\rightarrow$ PO+H), enhancing the formation of PO, but the fast depletion of PH and PH$_2$ affects PO negatively  because reactions 3 and 4 need PH and PH$_2$ to create PO. The combination of both effects makes PO to reach lower maximum abundances than for $T$=10 K, and consequently reaction 2 transforms PO into PN at a lower rate and hinders PO from reaching very low values. For this reason, PN becomes more abundant than PO (i.e. [PO]/[PN]$<1$) later than for $T$=10 K.

Interestingly, the numerical solutions of the total system and the theoretical solutions of the minimal system are indistinguishable to the naked eye in Figure~\ref{fig:numerical_evolution} for all chemical species, temperatures, and at all times, proving the suitability of the simplifications assumed to obtain the theoretical expressions in Equation~(\ref{eq:general_solution}). We compared the final abundances of the P-bearing species calculated via numerical methods with the same quantities obtained from our theoretical solution (Equation~(\ref{eq:general_solution})). We report average relative errors of $\sim0.3\%$ for $T$=10 K, $\sim1\%$ for $T$=100 K, and $\sim2\%$ for $T$=300 K. In summary, the mathematical solutions of the minimal system provide a highly accurate description of the evolution of the abundances of the P-bearing chemical species. Also, let us remark that the calculation of the final abundance (i.e. at time $t=10^5$ yrs) of a species with Equation~(\ref{eq:general_solution}) in a typical laptop computer is on average more than $10^5$ times faster than the numerical solution of the total system. 
 
\begin{figure*}[ht]
\centering
\includegraphics[width=\textwidth]{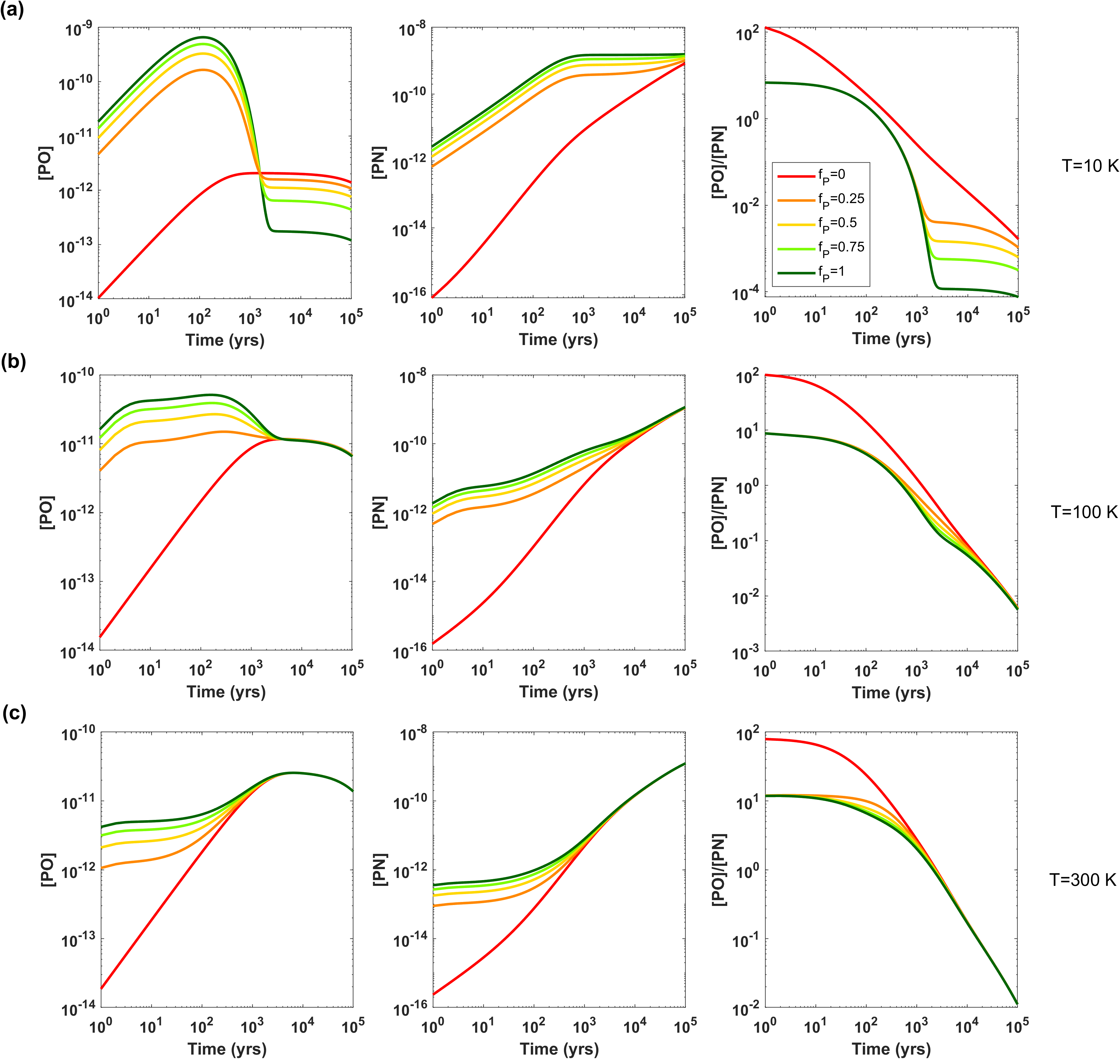}
\caption{Evolution of the abundance of PO, PN and their ratio [PO]/[PN] for (a) $T$=10 K, (b) $T$=100 K and (c) $T$=300 K and for different values of the P-hydrogenation fraction $f_\mathrm{P}$, the fraction of P that has been transformed into PH, PH$_2$ and PH$_3$ via grain-surface reactions before being released to the gas phase. }
\label{fig:fPt}
\end{figure*}

Finally, the minimal and total systems naturally yield that, at typical cloud ages (between $t=10^4$ and $t=10^5$ yrs) the abundance of PN is clearly larger than the abundance of PO, as predicted by other models. In the analysis developed above, we have identified potential sources that could be contributing to the [PO]/[PN] disagreement between observations and models: models provide final ratios [PO]/[PN]$<$1 for all temperatures because, for large times (i) the PO formation routes are not significant anymore (since they depend on PH and PH$_2$, which get depleted rapidly), and (ii) the transformation of PO into PN governs the system. We will address in detail the possible sources of the [PO]/[PN] disagreement between observational data and models in the Discussion.

\subsection{The Role of Grain-surface Chemistry}

The chemical evolution of the P-bearing species is also affected by the grain-surface reactions taking place in the physico-chemical environment where the system evolves. To evaluate this effect, we now focus on the dependence of the system on the hydrogenation fraction of P ($f_\mathrm{P}$), that represents
the fraction of P that is hydrogenated on dust grains via grain-surface reactions before being released to the gas phase. As introduced in Section$\,$\ref{theoreticalframework} and 
Table~\ref{tab:initial_abundances}, the initial abundance of P is given by $(1-f_\mathrm{P})\times 2.57 \times 10^{-9}$. For simplicity we assume that PH, PH$_2$ and PH$_3$ have equal initial abundances of $(f_\mathrm{P}/ 3) \times 2.57 \times 10^{-9}$. Unbalancing the initial abundances of PH, PH$_2$ and PH$_3$ would only have a noticeable effect for low temperatures (see Appendix~\ref{appendix:PHs}), but note that PO and PN have so far been reported only in star-forming regions affected by shocks \citep{cernicharo06,Bernal_2021,zeng18,Rivilla_2022, Lefloch_2016}, where the temperature is around 100 K or larger.

Figure~\ref{fig:fPt} shows the time-evolution of the abundances of PO, PN and their ratio for $T$=10 K, 100 K and 300 K, for different values of $f_\mathrm{P}$ from 0 to 1 (the dependence of the initial abundances on $f_\mathrm{P}$ is shown in Table~\ref{tab:initial_abundances}). 
We can see that PO at $T$=10 K presents a complex dependence on $f_\mathrm{P}$ because, as we explained in Section~\ref{subsec:numerical}, PO abundance does not decay to zero due to reaction 6 (P+OH $\rightarrow$ PO+H), which keeps its abundance above a certain limit. Since reaction 6 has P as a reactant, it is straightforward to see that this PO abundance limit for large times depends negatively on $f_\mathrm{P}$.

On the contrary, the abundances of PO and PN (and thus the ratio [PO]/[PN]) do not depend on $f_\mathrm{P}$ at the end of the cloud's evolution for $T$=100 K and $T$=300 K: $f_\mathrm{P}$ determines the initial abundance of PO and PN, but eventually the curves converge. This means that it is irrelevant whether the source of P is atomic P or the set of PH, PH$_2$ and PH$_3$ (or a combination of them), as the same amount of P will be finally transformed into PO and PN. Furthermore, and as long as $f_\mathrm{P}$ is not zero, for $T$=100 K and $T$=300 K the ratio [PO]/[PN] can be considered independent of $f_\mathrm{P}$ for all times. This proves that [PO]/[PN] is a more robust quantity for all times than [PO] and [PN] separately in order to compare observed data with numerical predictions obtained from existing models.

\subsection{Sensitivity of the Abundances of PO and PN on the Reaction Rate Coefficients}\label{subsec:governing}

\begin{figure*}[t]
\includegraphics[width=\textwidth]{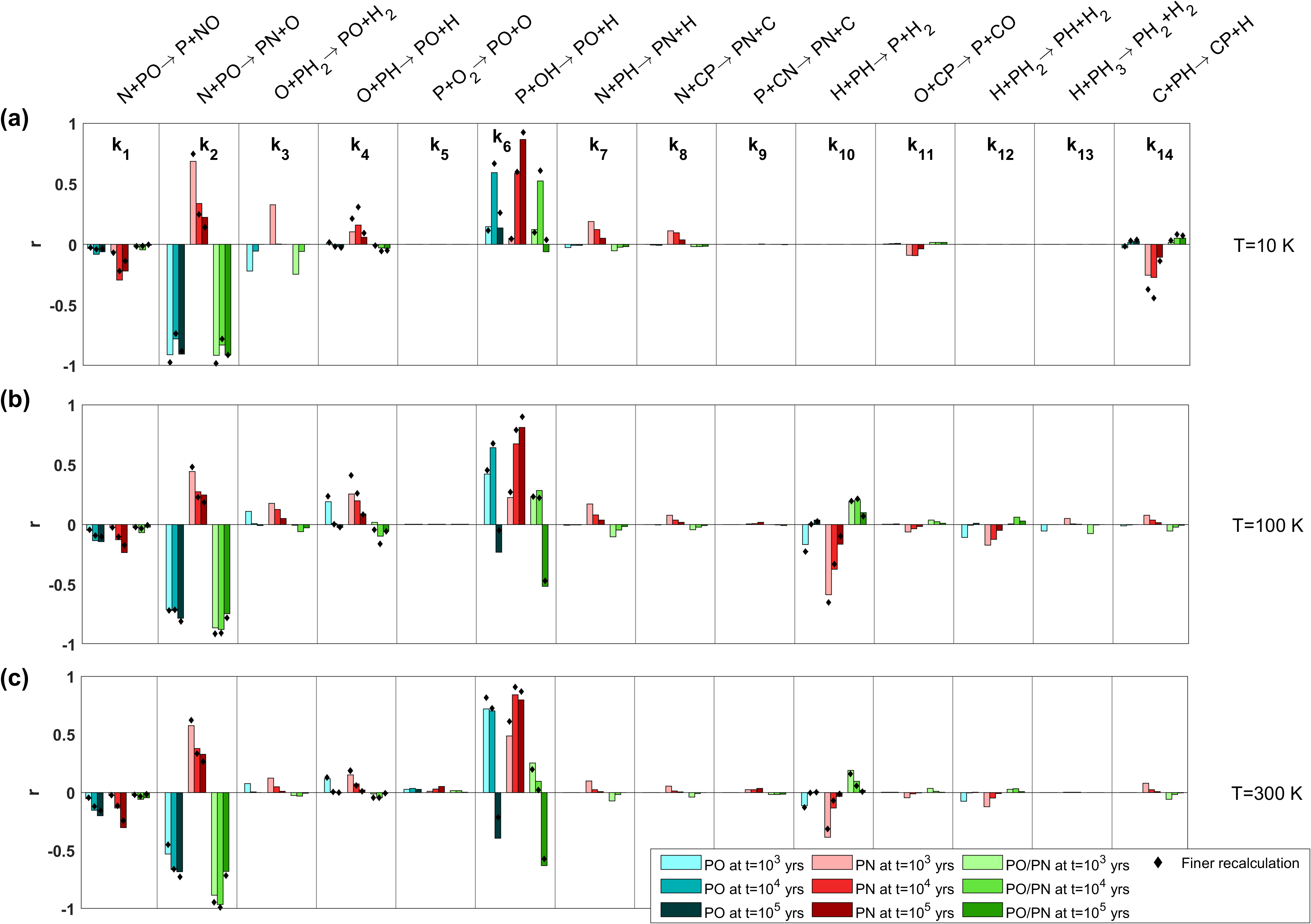}
\caption{Dependence of the formation of PO, PN and their ratio with the reaction rate coefficients $k_i$ of the 14 chemical reactions of the system. We plot the Pearson correlation $r$ of the abundance of PO (blue bars), PN (red bars) and their ratio (green bars)  with all 14 reaction rate coefficients for (a) $T$=10 K (b) $T$=100 K and (c) $T$=300 K, considering times $t=10^3$ (light colors), $t=10^4$ (vivid colors) and $t=10^5$ yrs (dark colors). Correlations $|r|>0.05$ are supported by p-values $<0.01$. The correlations were calculated with sets consisting of all the combinations of 3 values of each of the 14 $k_i$ (bars), and also for a finer grid composed of 11 values of each of the 5 most influential $k_i$ (black diamonds). In all cases $k_i$ ranges from $k_i/10$ to $10k_i$, being the values logarithmically distributed, and P-hydrogenation fraction $f_\mathrm{P}=0.5$. 
\label{fig:correlations}
} 
\end{figure*}

Many rate coefficients associated with the different chemical reactions that take place in astrophysical environments are either totally unknown or very uncertain \citep{McElroy13,Wakelam_2012, wakelam15}. The reactions involved in our model are not an exception, 
as the error associated with most of the reaction rate coefficients is at least 2-fold \citep{Wakelam_2012}. P is naturally highly reactive, and so treating it experimentally to obtain the kinetic parameters becomes especially challenging. It is also possible to apply theoretical quantum chemical methods for this purpose, but they are computationally very expensive and therefore it is not possible to apply them to all the reactions. Regarding our network, only reactions 5, 6 and 7 in Table~\ref{tab:set_of_reactions} have been calculated through these methods \citep{GarciadelaConcepcion_2021, GarciadelaConcepcion_2023, Gomes_2023}. 

In this section we investigate how the uncertainty associated with each chemical reaction affects the final abundances of PO and PN. In particular, we aim to identify which rate coefficients $k_i$ should be preferentially constrained via precise theoretical quantum calculations or measured experimentally in the laboratory, since increasing their certainty would yield better predictions of PO and PN abundances in current and future numerical modelling. 
We thus make use of the theoretical solution of the minimal system, whose extremely fast calculation permits us to explore the parameter space in a way that would be impossible to tackle numerically. Benefiting from this fact, we calculate with Equation~(\ref{eq:general_solution}) the abundance of PO, PN and their ratio for all the combinations of 3 different values of each $k_i$ ($k_i/10$, $k_i$ and $10k_i$, $k_i$ obtained from the source provided in Table~\ref{tab:set_of_reactions}) for all 14 reaction rate coefficients of the total system. In this way, we obtain sets of $3^{14}$ data for [PO], [PN] and [PO]/[PN], and do so for three different times: $t=10^3$, $t=10^4$ and $t=10^5$ yrs.
Figure~\ref{fig:correlations} shows the Pearson correlation coefficient $r$ between these sets of values of [PO] (blue bars), [PN] (red bars) and [PO]/[PN] (green bars) and each reaction rate coefficient $k_i$ (calculated in a log-log scale), for $T$=10 K, 100 K and 300 K. Bar colors go from light to dark according to the evolution times. 

Furthermore, to check that the correlations plotted in Figure~\ref{fig:correlations} are sufficiently precise in spite of the fact that we used only 3 values for each $k_i$ to save computer time, we also plot (in black diamonds) the Pearson coefficients calculated for a much finer grid of 11 different values for the 5 most influential $k_i$ (from $k_i/10$ to $10k_i$, including $k_i$, in a logarithmically uniform distribution), giving rise to $11^5$ different data for each set of [PO], [PN] and their ratio. The similarity between the correlations calculated with 3 and 11 different values is clear.

The results plotted in Figure~\ref{fig:correlations} show that, for $T$=10 K, the abundances of PO, PN and their ratio have a strong dependence on the reaction rate coefficients $k_2$ and $k_6$, while $k_1$, $k_3$ and $k_4$ and $k_{14}$ also play a significant role. The system's dependence on reaction rate coefficients is very similar for $T$=100 K and $T$=300 K, and in comparison to $T$=10 K, the abundances weaken their correlation with $k_{14}$ while $k_{10}$ becomes relevant because reaction 10 and its associated reactions (reactions 12 and 13) have a non-zero activation barrier ($\gamma$) that makes them highly dependent on temperature (see Table~\ref{tab:set_of_reactions} and Equation~(\ref{eq:arrhenius})). 

Analyzing Figure~\ref{fig:correlations} in  more detail, we find that, for all temperatures and times, the rate coefficient of reaction 2 (N+PO $\rightarrow$ PN+O) has a strong negative correlation with PO, and a strong positive correlation with PN (and thus the correlation with [PO]/[PN] is negative), as expected. 
The influence of reaction 6 (P+OH $\rightarrow$ PO+H) on the system, on the contrary, is more intricate because the correlations between its rate coefficient $k_{6}$ and the abundances strongly vary with time. It happens that $k_{6}$ is strongly positively correlated with the abundance of PN for all temperatures even when PN is not in reaction 6, and grows with time. This behavior relies on the fact that when time grows PN is mostly obtained from PO through reaction 2, explaining why the correlation is higher at longer times.
In a similar way, for all temperatures and short times ($t=10^3$ and $t=10^4$ yrs), the abundance of PO is positively correlated with $k_{6}$, as expected. However, for large temperatures and times the abundance of PO negatively correlates with $k_{6}$, and this apparently paradoxical effect can be explained as follows: increasing $k_{6}$ for large T accelerates the production of PO and the consumption of P, being beneficial for the growth of PO at short times, but, at the end of the cloud's evolution time ($t\sim10^5$ yrs), less P will be available to form PO and the system will not be able to counteract the negative effect of reaction 2 consuming PO. Consequently, enhancing a reaction that has PO as a product can lead to negative effects on it at certain physico-chemical conditions. With this example, we remark on the complexity of analyzing astrochemical models: even for our simple case (made of 14 chemical reactions), including formation routes for a certain chemical species may not result in an increase of that species abundance.

\section{Improving the certainty of the reaction rate coefficients through Bayesian statistics}\label{sec:bayes}

\begin{figure*}[ht]
\centering
\includegraphics[width=\textwidth]{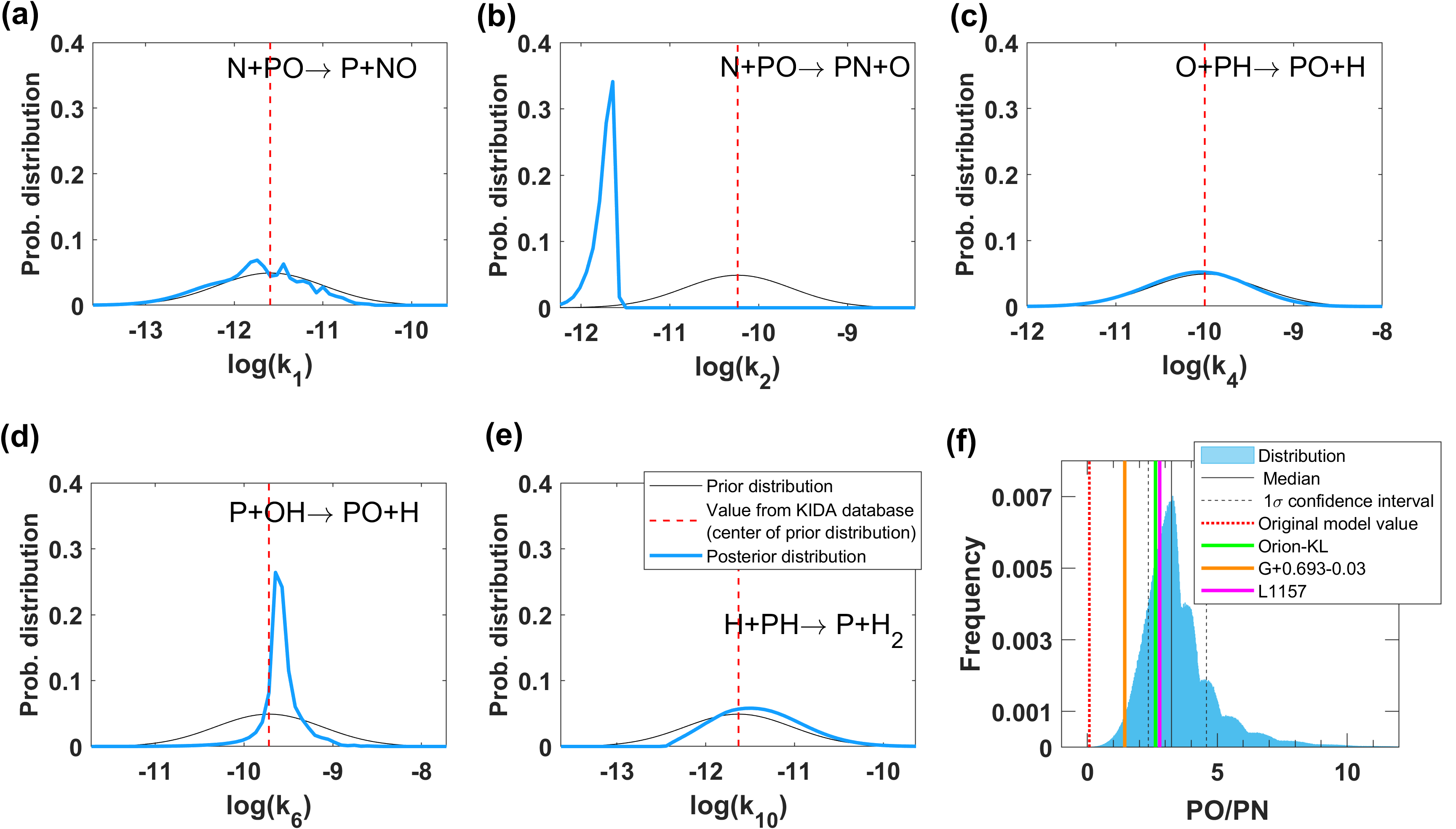}
\caption{Bayesian inference applied to the most important reaction rate coefficients of the model and the [PO]/[PN] ratio. (a-e) Prior probability distributions (thin black lines) and posterior probability distributions (PPDs, wide blue lines) obtained with Bayesian inference of the 5 most relevant reaction rate coefficients of our model for $T$=100 K and $t=10^{4}$ yrs, according to observations of star-forming regions from Table~\ref{tab:observations_bayes}. The center of the prior distribution is the value provided by KIDA and shown in Table~\ref{tab:set_of_reactions} (dashed red lines). (f) Distribution of the [PO]/[PN] abundance ratio obtained from sampling the PPD's of the reaction rate coefficients. The median (black line) along with its 1$\sigma$ confidence interval (dashed black lines) are shown, as well as the original abundances obtained in the model with KIDA values of the rate coefficients (dotted red line), and the real abundances from clouds Orion-KL, G+0.693-0.03 and L1157 (green, orange and magenta lines, respectively). P-hydrogenation fraction is $f_\mathrm{P}$=0.5 in all calculations.}
\label{bayes}
 
\end{figure*}

Bayesian statistics relies on the combination of available real data and a previous knowledge of the parameters of the system under study, and has gained significant popularity in the past decades. It has been applied to many different fields, including chemical kinetics \citep{Hsu_2009, Galagali_2015, Cohen_2021}, and recently proved to unveil relevant information about the model parameters and their associated uncertainties in the context of an astrochemical system \citep{Holdship_2018,Heyl_2020}. 

Following this methodology, in this section we apply Bayesian statistics to the evolution of ISM phosphorus with the aim of improving our knowledge of the reaction rate coefficients. The real data used correspond to observations of PO and PN abundances in the star forming region Orion-KL, the Giant Molecular Cloud G+0.693-0.03 located in the Galactic Center and the star-forming region L1157 (see Table~\ref{tab:observations_bayes}). The chemistry observed in these three sources is dominated by shocks \citep{cernicharo06,Bernal_2021,zeng18,Rivilla_2022, Lefloch_2016}. We have selected these sources because they are the only ones for which the PO and PN abundances have been reported with their associated uncertainties. Note that we are assuming that our model for $T$=100 K describes the physico-chemical environment of these sources (i.e. gas affected by shocks), and consequently all the calculations in this section have been carried out considering $T$=100 K. For this reason, we chose to apply Bayesian inference to the five most influential reaction rate coefficients in PO and PN at $T$=100 K, $k_m$, where  \textbf{$m=\{1,2,4,6,10\}$}, as obtained in Section \ref{subsec:governing}. The P-hydrogenation fraction $f_\mathrm{P}$ has been established to be $f_\mathrm{P}=0.5$, as in Sections \ref{subsec:numerical} and \ref{subsec:governing}, and we have fixed $t=10^4$ yrs for all calculations according to the estimated age of the  shocked regions Orion-KL \citep[see e.g.][]{cernicharo06}, G+0.693-0.03 \citep{requena06} and L1157 \citep{Gueth_1996,Podio_2016}.

\begin{deluxetable*}{cccccc}\label{tab:observations_bayes}
\tablecaption{Observational data of PO and PN abundances used for Bayesian inference of the model's reaction rate coefficients.}
\tablehead{\colhead{$q$} & \colhead{Source} & \colhead{[PO]} & \colhead{[PN]}  & \colhead{[PO]/[PN]} & \colhead{Reference} 
} 
\startdata
1 & Orion-KL & $(1.6 \pm 0.1 )\times  10^{-10} $                       & $(6.1 \pm 0.6) \times  10^{-11}$  & $2.6 \pm 0.4 $ &   \cite{Bernal_2021} \\
2 & G+0.693-0.03 & $(5.9 \pm 2.2 )\times  10^{-11} $ & $(4.1 \pm 0.2 )\times  10^{-11}$ & $1.4 \pm 0.6 $ & \cite{Rivilla_2018} \\
3 & L1157 & $(2.5 \pm 0.4 )\times  10^{-9} $ & $(9.0 \pm 1.0 )\times  10^{-10}$ & $2.8 \pm 0.5 $ & \cite{Lefloch_2016} \\
\enddata
\end{deluxetable*}

The Bayes’s rule yields the posterior probability distributions (PPDs) of the parameters of the model (in this case the reaction rate coefficients) associated with any set of calculated abundances, and it is described as follows:

\begin{equation}\label{eq:bayes_posterior}
P( \mathbf{k}_j | \mathbf{x}) = \frac{P(\mathbf{x} | \mathbf{k}_j)  P(\mathbf{k}_j)}{\sum_j P(\mathbf{x} | \mathbf{k}_j)  P(\mathbf{k}_j)} \propto P(\mathbf{x} | \mathbf{k}_j)  P(\mathbf{k}_j)\,,
\end{equation}
where {\bf x} is the real data 
and $\mathbf{k}_j$ represents every set of values of the reaction rate coefficients. Our target is to obtain $P( \mathbf{k}_j |\mathbf{x})$, the posterior probability of each set, as it represents the certainty of the reaction rate coefficients after considering all the available data and any previous knowledge that we might have of the values of the parameters and their uncertainties. The denominator is
the sum of the probabilities of all the sets, that is, a normalization constant.
In summary, we need to calculate  $P(\mathbf{k}_j)$ and $P(\mathbf{x} | \mathbf{k}_j)$ for every $\mathbf{k}_j$ to obtain the PPDs associated with the reaction rate coefficients.

$P(\mathbf{k}_j)$ is the prior probability of a given set of values of the reaction rate coefficients $\mathbf{k}_j$. To calculate it, Bayesian methodology requires the definition of a prior probability distribution $P (k_m)$ for each of the five reaction rate coefficients $k_m$. 
Following the information in KIDA \citep{Wakelam_2012}, 
we established a discrete 55-value log-normal distribution centered in the $k_m$ provided in Table~\ref{tab:set_of_reactions}, in a range of $k_{m,l}$ with $l=\{1,...,55\}$ from $k_{m,1}=k_m/100$ to $k_{m,55}=100k_m$, and with a standard deviation established in a way that $4k_m$ and $1/4 k_m$ fall inside the 68.2 \% confidence interval ($1\sigma$). The choice of 55 points for each rate coefficient $\mathbf{k}_j$ ensures a sufficiently exhaustive analysis with moderate computational costs. Appendix~\ref{appendix:prior} analyzes the same system considering a log-uniform prior probability distribution. 

$\mathbf{k}_j=\{k_1,k_2,k_4,k_6,k_{10}\}$ describes every set of 5 values of the reaction rate coefficients $k_m$ chosen in the range $k_{m,l}$, where   $j$ therefore stands for the $55^5$ different combinations of $l=1,...,55$ values for each of the 5 reaction rate coefficients.
In consequence, since the value of each rate coefficient is independent, the joint probability of the set $\mathbf{k}_j$ is
\begin{equation}\label{eq:joint}
\centering
P(\mathbf{k}_j)=\prod_m^5 P (k_{m,l})\,.
\end{equation}
$P(\mathbf{x} | \mathbf{k}_j)$ is the likelihood of $\mathbf{k}_j$, i.e. the probability of obtaining from the model the observed values of PO and PN given a specific set $\mathbf{k}_j$, and it is obtained as 
\begin{equation}
    P(\mathbf{x}| \mathbf{k}_j)= \prod_q^3 \prod_r^2 \mathrm{exp} \left( -\frac{1}{2} \left( \frac{x_{r,q}-[X_{r,j}]}{\sigma_{r,q}} \right)^2 \right)\, ,\label{eq:likelihood}
\end{equation}
where $q=\{1,2,3\}$ refers to each observational source (see Table~\ref{tab:observations_bayes}) and $r=$\{PO,PN\}, in such a way that $x_{r,q}$ represents the observed abundance of species $r$ in source $q$ (along with its standard deviation $\sigma_{r,q}$) and  $[X_{r,j}]$ is the model's predicted final abundance of species $r$ for a given set $\mathbf{k}_j$. 
Calculating $[X_{r,j}]$ for all the $55^5$ combinations of reaction rate coefficients was computationally accessible once again because of the theoretical solution of the system introduced in Equation~(\ref{eq:general_solution}). As we could explore the totality of the parameter space, we avoided the usual necessity of a much more complex Markov Chain Monte Carlo (MCMC) sampling method \citep{Holdship_2018}.

Once we have calculated $P(\mathbf{x} | \mathbf{k}_j)$ for all $\mathbf{k}_j$ following Equations~(\ref{eq:bayes_posterior}-\ref{eq:likelihood}), we can finally obtain the PPD of a given $k_m$. To do so, we need to sum up all the probabilities $P(\mathbf{x}| \mathbf{k}_j)$ for which the set $\mathbf{k}_j$ contains a fixed $k_m$ at its value $k_{m,l}$, as follows
\begin{equation}\label{eq:marginal}
    P(k_m = k_{m,l} | \mathbf{x}) =\sum_{j/k_{m}=k_{m,l}} P( \mathbf{k}_j | \mathbf{x})\,.
\end{equation} 
Note that the sum contains $55^4$ elements for which the values of the fixed $k_m$ are equal to $k_{m,l}$ and the values of the other $k_m$ take all their 55 possible values. Then, the PPD of $k_m$ is given by the results of Equation~(\ref{eq:marginal}) for the 55 values of $k_{m,l}$.
The prior probability distributions $P(k_m)$ and the PPDs for each $k_m$ are plotted in Figure~\ref{bayes}(a-e). If we compare the peak of each posterior probability distribution with the original value provided by KIDA (red dashed vertical line), we can make estimates of the reaction rate coefficients according to the observations of PO and PN that we considered. 

Figure~\ref{bayes}(a-e) shows that the PPDs of $k_2$ and $k_6$
 present much higher and sharper peaks than the distributions of the other reaction rate coefficients, in agreement with what we already obtained in Section~\ref{subsec:governing}: the abundances of PO and PN depend critically on the values of $k_2$ and $k_6$ at this time and temperature. On the one hand, $k_2$ presents a peak at a value that is $\sim0.04$ times the value in KIDA. This reveals that the available value of $k_2$ might be a large overestimation of the real one.
On the other hand, $k_6$ PPD is centered very close to its available value in agreement with the fact that $k_6$ was calculated with precise theoretical methods \citep{GarciadelaConcepcion_2021}. 
The rest of the rate coefficients ($k_1$, $k_4$ and $k_{10}$) do not have a strong impact on PO and PN abundances, as Figure~\ref{bayes}(a,c,e) PPDs are similar to the prior distributions assigned to them. However, if we calculate the posterior probability distributions of the system making use of a log-uniform prior instead of a log-normal prior, that is, if we use priors devoid of information, the results plotted in Appendix~\ref{appendix:prior} confirm that the observational data do not provide any relevant information about $k_4$, 
but allow us to constrain the values of $k_1$ and $k_{10}$ so that $k_1<3.9\times 10^{-11}$cm$^{3}$ s$^{-1}$ and $k_{10}>3.0\times 10^{-13}$cm$^{3}$ s$^{-1}$. As shown in Appendix~\ref{appendix:prior}, $k_{6}$ is constrained to the value calculated by \citet{GarciadelaConcepcion_2021} even when using a log-uniform prior.

Finally, we sampled the posterior probability distributions for the abundance of PO, PN and their ratio at $t=10^4$ yrs and $T$=100 K, and plotted the latter in Figure~\ref{bayes}(f). To do so, we used the theoretical solution of the system (given by Equation~(\ref{eq:general_solution})). We found that the median of PO and PN abundances are around one order of magnitude larger than the original numerical values (see Table~\ref{tab:results_bayes}), and the median of [PO]/[PN] is almost two orders of magnitude larger than the original numerical value (see the dotted red line in Figure~\ref{bayes}(f) and Table~\ref{tab:results_bayes}), in agreement with observations.

\begin{deluxetable}{cccc}\label{tab:results_bayes}
\tablecaption{Model's outputs before and after applying Bayesian inference to the reaction rate coefficients.}
\tablehead{\colhead{Data source} & \colhead{[PO]} & \colhead{[PN]}  & \colhead{[PO]/[PN]} 
}
\startdata
Original   & $1.12\times  10^{-11} $                       & $1.67 \times  10^{-10}$  &      0.067 \\
Bayesian inferred & $1.4\times  10^{-10} $ & $4.2\times  10^{-11} $ & 3.3 \\
\enddata
\tablecomments{The first line shows the abundances of PO, PN and their ratio calculated applying the original (that is, without Bayesian inference) values of the reaction rate coefficients compiled in Table~\ref{tab:set_of_reactions}. The second line shows the median of the distributions of the abundances of PO, PN and their ratio calculated from sampling the posterior probability distributions of reaction rate coefficients $k_m$ (where  $m=\{1,2,4,6,10\}$), and the original values for the rest. In both cases the abundances have been calculated with the theoretical model (Equation~(\ref{eq:general_solution})) applying $T$=100 K, P-hydrogenation fraction $f_\mathrm{P}$=0.5 and $t=10^4$ yrs.}
\end{deluxetable}

\section{Discussion}

In this work, we have developed a thorough theoretical and numerical study of the dynamical system associated with the chemical evolution of phosphorus in an interstellar molecular cloud. A wide variety of techniques and algorithms have been developed for network reduction in chemical models \citep{Tupper_2002, Lehmann_2004, Markosyan_2014, Peerenboom_2015, Ayilaran_2019}, some of them focusing precisely on astrochemical systems \citep{Hollenbach_2009,Heyl_2020}. Here we present a different approach.
By making suitable assumptions, we have focused on a complex and limited network of phosphorus with 14 chemical reactions and 17 chemical species. This system can be further reduced so that it only analyzes the evolution of the P-bearing species PO, PN, CP, P, PH, PH$_2$ and PH$_3$, becoming a solvable system made of 7 linear ODEs. The grain-surface chemistry is taken into account in the model through the parameter $f_\mathrm{P}$, the fraction of P that has been transformed into PH, PH$_2$ and PH$_3$ before being released to the gas phase.

Most studies in recent literature that model the evolution of the chemistry of phosphorus in the interstellar medium and in star-forming regions ground on the use of complex software describing the specific physico-chemical conditions of the target astronomical source. These computer programs consider some thousands of chemical reactions with uncertain rate constants. In contrast, we have focused exclusively on the most relevant reactions regarding the phosphorus chemistry in order to approach the system from a theoretical perspective and benefit from a much deeper knowledge of its complex dynamics. In particular, the selected reactions are neutral-neutral reactions, which in general lack accurate laboratory measurements and which are dominant in the interstellar regions where P-bearing species are detected (as e.g. in regions dominated by shocks).
Furthermore, the explicit mathematical expressions obtained for the chemical evolution of the relevant species allowed us to develop a thorough analysis of the phenomenology with computation times that were up to five orders of magnitude faster than the numerical methods needed to solve the total system, and obviously millions of times faster than with the use of any complex astrochemical software. 

We have detected several target reactions whose rate coefficients should be determined accurately in future calculations or experiments in order to minimize the uncertainty in the astrochemistry of phosphorus.
The evolution of the P chemical network is sensitive to a reduced set of key reactions for low temperatures (leading the conversion of PO into PN by N+PO $\rightarrow$ PN+O --reaction 2-- or the conversion of PH and PH$_2$ into PO and PN through O+PH$_2$ $\rightarrow$ PO+H$_2$ --reaction 3--, O+PH $\rightarrow$ PO+H --reaction 4-- and N+PH $\rightarrow$ PN+H --reaction 7--, while for high temperatures the chemistry becomes more complex and involves these and other interactions, such as the intensive destruction of PH in H+PH $\rightarrow$ P+H$_2$ --reaction 10--.  
Furthermore, Bayesian methods applied to the model at $T$=100 K and the use of real data regarding 3 different sources (Orion-KL, G+0.693-0.03 and L1157) yield that the reaction rate coefficient $k_2$ might be especially overestimated, according to our results by a factor of $\sim$25. 

Unveiling the formation dynamics of PO and PN over time helped us to identify possible sources of the [PO]/[PN] disagreement between observational data and models. Observational data yield [PO]/[PN] $\sim1.4-3$ in star forming regions \citep{Ziurys_1987,Fontani16, Rivilla_2016, Lefloch_2016, Rivilla_2018, Rivilla_2020, Bernal_2021, Bergner19, Bergner22}, with the exception of [PO]/[PN]$=0.6\pm0.5$ recently detected toward Ser SMM1, which is subjected to large uncertainties \citep{Wurmser_2022}.  
In contrast, numerical models typically yield [PO]/[PN]$<$1 \citep{Jimenez-Serra_2018,Chantzos_2020,Sil_2021}. Our simulations show that [PO]/[PN] grows with the temperature of the cloud, but still yields values of [PO]/[PN] $<<1$ for all the scenarios analyzed with the parameters present in Table~\ref{tab:set_of_reactions} and the initial conditions in Table~\ref{tab:initial_abundances} for times between $t=10^4$ and $10^5$ yrs because, at the final stages of the evolution, the formation routes of PO become negligible while reaction 2 (N+PO $\rightarrow$ PN+O) governs the system. Therefore, we argue that current astrochemical models are unable to yield realistic [PO]/[PN] values because (i) certain reaction rate coefficients, mainly $k_2$, are estimated very inaccurately, and (ii) models might lack important destruction routes for PN (e.g. in KIDA \citep{Wakelam_2012} only N+PN $\rightarrow$ P+N$_2$ is present and its kinetic parameters $\alpha=10^{-18}$ cm$^{3}$ s$^{-1}$ and $\beta=\gamma=0$ make it negligible in molecular clouds).  
We stress that we have not considered either ion-neutral reactions or photochemistry in this work. However, note that these are valid assumptions given that most regions where PO and PN have been detected present a chemistry dominated by shocks and not by photochemistry. Indeed, when compared to the astrochemical code UCLCHEM, our model reproduces well the evolution of the [PO]/[PN] ratio with time for the same physical conditions and initial abundances (see Figure~\ref{fig:numerical_evolution}(d-f)).

Interestingly, [PO]/[PN] is not dependent on the P-hydrogenation fraction $f_\mathrm{P}$ for any time of the evolution at high temperatures ($T$=100 K and 300 K), as long as a small amount of P is converted into PH$_3$, that is, for $f_\mathrm{P}>0$. This reinforces the prevalence of [PO]/[PN] over [PO] and [PN] separately in order to compare observed data with numerical predictions obtained from existing models.

In spite of the already mentioned predominance of PN over PO at the final stages of the cloud evolution, our environment shows a natural prevalence of PO over PN at the very beginning of the evolution of the system (even taking into account that [PO]$_0$=[PN]$_0$=0). The theoretical solution of the system allows for a calculation of the limit of [PO]/[PN] at early times, which gives 
   \begin{equation}\label{eq:limitePOPN}
     \lim_{t \to 0} \frac{[PO]}{[PN]}\approx \frac{(k_{3}+k_{4})[\mathrm{O}]_0}{k_{7}[\mathrm{\mathrm{N}}]_0}\,,
  \end{equation}
(see Appendix~\ref{appendix:ratio} for the mathematical proof). Note that this expression does not depend on 
the P-hydrogenation fraction $f_{\mathrm{P}}$ (as long as $f_{\mathrm{P}}>0.01$ to ensure the existence of sufficient initial PH and PH$_2$, condition required to perform the approximations in Appendix~\ref{appendix:ratio}) or the initial abundances with the exception of N and O. If we evaluate Equation~(\ref{eq:limitePOPN}), we obtain that [PO]/[PN] ranges from 7 to 12 depending on the temperature, which agrees with numerical results at early times for all T and $f_{\mathrm{P}}$ with an average error of $0.5\%$. Although this result is limited by the model's caveats, we can extract some general conclusions: at early times, only reactions O+PH$_2$ $\rightarrow$ PO+H$_2$, O+PH $\rightarrow$ PO+H and N+PH $\rightarrow$ PN+H --reactions 3, 4 and 7 respectively-- are relevant, and PO formation seems to be much more enhanced than PN formation because (i) the cosmic abundance of O is one order of magnitude higher than the cosmic abundance of N; and (ii) the addition of reaction rate coefficients $k_3$ and $k_4$ is similar to $k_7$ at all T. However, the ratio decreases when at longer times reaction N+PO $\rightarrow$ PN+O --reaction 2- becomes noticeable and reinforces PN, which overcomes PO eventually leading to the [PO]/[PN] values under 1 typically obtained from models at $t\sim10^4-10^5$ yrs. In fact, the theoretical solution for PO and PN (Equation~(\ref{eq:general_solution})) yields that [PO]$\rightarrow 0$ and [PN]$\rightarrow C_{77}>0$ when $t\rightarrow\infty$, which means that [PN] will sooner or later overcome [PO] leading to [PO]/[PN]$<1$ for any value of the reaction rate coefficients and initial conditions. However, let us remark that, when the corrected values of the reaction rate coefficients obtained by Bayesian inference are included (in particular when $k_2$ decreases), the crossing-time between PO and PN grows and PO remains more abundant than PN for meaningful evolution times (i.e. $t\sim 10^4$ yrs or more), solving in this way the [PO]/[PN] disagreement between models and real data. 
 
Finally, we believe that the analysis of astrochemical systems with the tools of network theory is a promising line of research that has not been sufficiently developed yet. While the goal of the seminal studies linking both disciplines was the topological description of astrochemical networks associated with very diverse astrophysical environments \citep{Sole_2004, Jolley_2010}, more recently the geometry of grain surface reaction networks was analyzed to reduce the computational expense of performing Bayesian inference \citep{Heyl_2020}, and the emergence of interstellar molecular complexity was successfully explained with a model based on interacting complex networks \citep{Garcia_2022}.
We are confident that our multidisciplinar approach will attract the attention of both the astrochemistry and complexity theory communities in the next years, as representing astrochemical systems as complex networks in permanent evolution provides a profound understanding of the formation and destruction of the chemical species involved. Larger or more complex chemical networks than the phosphorus network might preclude the calculation of an explicit theoretical solution and its concomitant drastic decrease in computer time, but with more computational power the methodology here introduced could still be used to describe the formation of a wide variety of chemical precursors of organic macromolecules in space, a key question to unveil the origin and early evolution of life on Earth. 

\begin{acknowledgments}
The authors acknowledge insightful comments on the manuscript from S. Viti, technical advice on Bayesian Inference from M. Castro, and fruitful conversations with A. Aguirre-Tamaral, J. Garc\'{\i}a de la Concepci\'on, R. Guantes, S. Manrubia, A. Meg\'{\i}as, V.M. Rivilla and M. Ruiz-Bermejo. 
J.A. and M.F.-R. received support from grant No. PID2021-122936NB-I00, J.A. and I.J.-S. from grant No. PID2019-105552RB-C41 and J.A., M.F.-R. and I.J.-S. from grant No. MDM-2017-0737 Unidad de Excelencia ``Mar\'ia de Maeztu''-Centro de Astrobiolog\'ia (CSIC-INTA), funded by the Spanish Ministry of Science and Innovation/State Agency of Research MCIN/AEI/10.13039/501100011033 and by ``ERDF A way of making Europe''.   
\end{acknowledgments}

\appendix
\section{Set of ordinary differential equations that describes the chemical evolution of phosphorus in the interstellar medium}\label{appendix:ODEs}

The dynamical system under study is a set of 14 reactions which involve 17 chemical species. Applying the law of mass action to the reactions, we obtain a system of 17 ordinary differential equations (ODEs) which accounts for the evolution with time of each chemical species abundance. The system is:
\begin{eqnarray}\label{eq:total_system_begin}
 \frac{1}{n_{\mathrm{H}}}       \frac{d[\mathrm{C}]}{dt}   &=& k_{8}[\mathrm{N}][\mathrm{CP}]+k_{9}[\mathrm{P}][\mathrm{CN}]-k_{14}[\mathrm{C}][\mathrm{PH}] \\ 
     \frac{1}{n_{\mathrm{H}}}     \frac{d[\mathrm{CN}]}{dt}   &=& -k_{9}[\mathrm{P}][\mathrm{CN}]\\
 \frac{1}{n_{\mathrm{H}}} \frac{d[\mathrm{CO}]}{dt}   &=&k_{11}[\mathrm{O}][\mathrm{CP}]\\
 \frac{1}{n_{\mathrm{H}}} \frac{d[\mathrm{CP}]}{dt}   &=&-k_{8}[\mathrm{\mathrm{N}}][\mathrm{CP}]-k_{11}[\mathrm{O}][\mathrm{CP}]+k_{14}[\mathrm{C}][\mathrm{PH}]\\
 \frac{1}{n_{\mathrm{H}}} \frac{d[\mathrm{H}]}{dt}   &=&k_{4}[\mathrm{O}][\mathrm{PH}]+k_{6}[\mathrm{P}][\mathrm{O}\mathrm{H}]+k_{7}[\mathrm{\mathrm{N}}][\mathrm{PH}]-k_{10}[\mathrm{H}][\mathrm{PH}]-k_{12}[\mathrm{H}][\mathrm{PH}_2]-k_{13}[\mathrm{H}][\mathrm{PH}_3]+k_{14}[\mathrm{C}][\mathrm{PH}]\\
 \frac{1}{n_{\mathrm{H}}} \frac{d[\mathrm{H}_2]}{dt}   &=&k_{3}[\mathrm{O}][\mathrm{PH}_2]+k_{10}[\mathrm{H}][\mathrm{PH}]+k_{12}[\mathrm{H}][\mathrm{PH}_2]+k_{13}[\mathrm{H}][\mathrm{PH}_3]\\
 \frac{1}{n_{\mathrm{H}}} \frac{d[\mathrm{N}]}{dt}   &=&-k_{1}[\mathrm{\mathrm{N}}][\mathrm{PO}]-k_{2}[\mathrm{\mathrm{N}}][\mathrm{PO}]-k_{7}[\mathrm{\mathrm{N}}][\mathrm{PH}]-k_{8}[\mathrm{CP}][\mathrm{\mathrm{N}}]\\
 \frac{1}{n_{\mathrm{H}}} \frac{d[\mathrm{N}\mathrm{O}]}{dt}   &=&k_{1}[\mathrm{N}][\mathrm{PO}]\\
 \frac{1}{n_{\mathrm{H}}} \frac{d[\mathrm{O}]}{dt}   &=&k_{2}[\mathrm{\mathrm{N}}][\mathrm{PO}]-k_{3}[\mathrm{O}][\mathrm{PH}_2]-k_{4}[\mathrm{O}][\mathrm{PH}]+k_{5}[\mathrm{P}][\mathrm{O}_2]-k_{11}[\mathrm{O}][\mathrm{CP}]\\
 \frac{1}{n_{\mathrm{H}}} \frac{d[\mathrm{O}_2]}{dt}   &=&-k_{5}[\mathrm{P}][\mathrm{O}_2]\\
 \frac{1}{n_{\mathrm{H}}} \frac{d[\mathrm{O}\mathrm{H}]}{dt}   &=&-k_{6}[\mathrm{P}][\mathrm{O}\mathrm{H}]\\
 \frac{1}{n_{\mathrm{H}}} \frac{d[\mathrm{P}]}{dt}   &=&k_{1}[\mathrm{\mathrm{N}}][\mathrm{PO}]-k_{5}[\mathrm{P}][\mathrm{O}_2]-k_{6}[\mathrm{P}][\mathrm{O}\mathrm{H}]-k_{9}[\mathrm{P}][\mathrm{CN}]+k_{10}[\mathrm{H}][\mathrm{PH}]+k_{11}[\mathrm{O}][\mathrm{CP}]\label{Eq:P}\\ 
 \frac{1}{n_{\mathrm{H}}} \frac{d[\mathrm{PH}]}{dt}   &=&-k_{4}[\mathrm{O}][\mathrm{PH}]-k_{7}[\mathrm{\mathrm{N}}][\mathrm{PH}]-k_{10}[\mathrm{H}][\mathrm{PH}]+k_{12}[\mathrm{H}][\mathrm{PH}_2]-k_{14}[\mathrm{C}][\mathrm{PH}]\\
 \frac{1}{n_{\mathrm{H}}} \frac{d[\mathrm{PH}_2]}{dt}   &=&-k_{3}[\mathrm{O}][\mathrm{PH}_2]-k_{12}[\mathrm{H}][\mathrm{PH}_2]+k_{13}[\mathrm{H}][\mathrm{PH}_3]\\
 \frac{1}{n_{\mathrm{H}}} \frac{d[\mathrm{PH}_3]}{dt}   &=&-k_{13}[\mathrm{H}][\mathrm{PH}_3]\\
 \frac{1}{n_{\mathrm{H}}} \frac{d[\mathrm{PN}]}{dt}   &=&k_{2}[\mathrm{\mathrm{N}}][\mathrm{PO}]+k_{7}[\mathrm{\mathrm{N}}][\mathrm{PH}]+k_{8}[\mathrm{\mathrm{N}}][\mathrm{CP}]+k_{9}[\mathrm{P}][\mathrm{CN}]\label{eq:total_system_PN}\\
 \frac{1}{n_{\mathrm{H}}} \frac{d[\mathrm{PO}]}{dt}   &=&-k_{1}[\mathrm{\mathrm{N}}][\mathrm{PO}]-k_{2}[\mathrm{\mathrm{N}}][\mathrm{PO}]+k_{3}[\mathrm{O}][\mathrm{PH}_2]+k_{4}[\mathrm{O}][\mathrm{PH}]+k_{5}[\mathrm{O}_2][\mathrm{P}]+k_{6}[\mathrm{O}\mathrm{H}][\mathrm{P}]\label{eq:total_system_end}
\end{eqnarray}

Note that this system of ODEs is a nonlinear system of the form $d\textbf{\textit{X}}/dt=\textbf{\textit{F}}(\textbf{\textit{X}})$, where $\textbf{\textit{X}}$ is the vector of the abundances of all chemical species. 

\section{Analytical solution of the system}\label{appendix:analytical}

The {\it total} system presented in Equations~(\ref{eq:total_system_begin}-\ref{eq:total_system_end}) is far too complex to be fully solved theoretically and must be treated numerically. However, in this Appendix we show that through pertinent approximations it can be linearized and simplified to obtain a {\it minimal} system that allows us to obtain explicit equations that fit very precisely the numerical evolution of the P-bearing species abundances.

Interestingly, all of kinetic Equations~(\ref{eq:total_system_begin}-\ref{eq:total_system_end}) are composed of a sum of terms which are in turn composed of a product of an \textit{abundant} species abundance and a \textit{scarce} species abundance, according to the classification explained in Section~\ref{theoreticalframework}. The only exception is the term $k_{9}[\mathrm{P}][\mathrm{CN}]$, in which P and CN species are both \textit{scarce}. 

In this work, we assume that the \textit{abundant} species abundances are constant (i.e. $d[X_i]/dt=0$ for C, H, N, O, O$_2$ and OH). Appendix~\ref{appendix:approximation} is devoted to describe the applicability and caveats of this assumption. In addition, we assume that the CN abundance is constant as its rate of change, $d[\mathrm{CN}]/dt = -k_{9} n_{\mathrm{H}}[\mathrm{P}][\mathrm{CN}]$, is extremely small because both P and CN are scarce species, and we neglect the term $k_1 [\mathrm{N}][\mathrm{PO}]$ in Equation~(\ref{Eq:P}) because its value is several orders of magnitude lower than the dominant terms.

The assumptions presented above convert the system described by Equations~(\ref{eq:total_system_begin}-\ref{eq:total_system_end}) in a linear system consisting of 10 ODEs. However, the non-interacting species NO, H$_2$ and CO do not influence the evolution of the rest of molecules (note that their abundances do not appear in the right-hand terms of the ODEs, a direct consequence of the fact that these species are not reactants of any reaction). Since we are only interested in the evolution of the P-bearing species, we can then neglect the kinetic equations relating NO, H$_2$ and CO and obtain that the total system described in Equations~(\ref{eq:total_system_begin}-\ref{eq:total_system_end}) can be finally reduced to a system of 7 ODEs composed of the equations of the rate of change of the abundances of PH$_3$, PH$_2$, PH, CP, P, PO and PN as
\begin{equation}
    \frac{d \textbf{\textit{X}}_\mathrm{P}}{dt}=\textbf{\textit{F}}_\mathrm{P}(\textbf{\textit{X}}_\mathrm{P})={\bf A} \textbf{\textit{X}}_\mathrm{P}\,,\label{Eq:reduced}
\end{equation}
where, if we number the species as in Table~\ref{species_numbers}, the vector containing the P-bearing species abundances is then $\textbf{\textit{X}}_\mathrm{P}$=\{[PH$_3$], [PH$_2$], [PH], [CP], [P], [PO] and [PN]\}, and consistently the matrix of coefficients {\bf A} becomes

$${\bf A}= \left( \matrix{ -r_1 & 0 & 0 & 0 & 0 & 0 & 0 \cr
k_{13} n_{\mathrm{H}} [\mathrm{H}] & -r_2 & 0 & 0 & 0 & 0 & 0 \cr
0 & k_{12} n_{\mathrm{H}} [\mathrm{H}] & -r_3 & 0 & 0 & 0 & 0 \cr
0 & 0 & k_{14} n_{\mathrm{H}} [\mathrm{C}] & -r_4 & 0 & 0 & 0 \cr
0 & 0 & k_{10} n_{\mathrm{H}} [\mathrm{H}] & k_{11} n_{\mathrm{H}} [\mathrm{O}] & -r_5 & 0 & 0 \cr
0 & k_{3} n_{\mathrm{H}} [\mathrm{O}] & k_{4} n_{\mathrm{H}} [\mathrm{O}] & 0 & k_{5} n_{\mathrm{H}} [\mathrm{O}_2]+k_{6} n_{\mathrm{H}} [\mathrm{O}\mathrm{H}] & -r_6 & 0 \cr
0 & 0 & k_{7} n_{\mathrm{H}} [\mathrm{N}] & k_{8} n_{\mathrm{H}} [\mathrm{N}] & k_{9} n_{\mathrm{H}} [\mathrm{CN}] & k_{2} n_{\mathrm{H}} [\mathrm{N}] & 0  \cr} \right)\,,$$
where constants $r_i$ are defined at the end of this Section. Note that throughout the paper we have named {\it minimal system} to the one defined by Equation~(\ref{Eq:reduced}).

\begin{deluxetable}{cccccccc}\tablecaption{P-bearing chemical species included in the minimal system (the one theoretically solved in this work). The indices $i$ determine the order in which the species abundances appear in vector $\textbf{\textit{X}}_\mathrm{P}$.\label{species_numbers}}
\tablehead{\colhead{Chemical species} & \colhead{PH$_3$} & \colhead{PH$_2$}  & \colhead{PH} & \colhead{CP} &\colhead{P} &\colhead{PO} &\colhead{PN}} 
\startdata
Index $i$  & 1         & 2 & 3 & 4 & 5 & 6 & 7 
\enddata
\end{deluxetable}
Solving this simplified system of ODEs is equivalent to obtaining the eigenvalues and eigenvectors of {\bf A}. Since {\bf A} is a matrix of size 7$\times$7, it would not be possible to solve its associated system of ODEs mathematically if it were not for the fact that {\bf A} is a triangular matrix. 
This configuration permits the resolution of each ODE sequentially. The calculations along with the solutions are presented below. Note that for simplicity the expressions obtained in this section for the different $[X_i](t)$ will be a function of the time $t$ and two sets of constants, $C_{ij}$ and $r_{i}$. The dependence of such constants on the initial conditions of the system (i.e. $[X_i](0)$) and the system parameters (i.e. the reaction rate coefficients $k_{i}$) is listed at the end of this Appendix.  \\

{\bf Calculation of [PH$_3$]:} The differential equation associated with PH$_3$ is
\begin{equation}\label{ph3_solucion}
   \frac{1}{n_{\mathrm{H}}} \frac{d[\mathrm{PH}_3]}{dt}=-k_{13}[\mathrm{H}][\mathrm{PH}_3]\,.
\end{equation}
Assuming $[\mathrm{H}]=[\mathrm{H}]_0$ (constant for all $t$), the only variable in Equation~(\ref{ph3_solucion}) is $[\mathrm{PH}_3]$, making it analytically solvable. We obtain
\begin{equation}\label{Sol:PH3}
    [\mathrm{PH}_3](t)=C_{11}\,e^{-r_1\,t}\,.
\end{equation}

{\bf Calculation of [PH$_2$]:} The differential equation associated with PH$_2$ is
\begin{equation}
   \frac{1}{n_{\mathrm{H}}} \frac{d[\mathrm{PH}_2]}{dt}=-k_{3}[\mathrm{O}][\mathrm{PH}_2]-k_{12}[\mathrm{H}][\mathrm{PH}_2]+k_{13}[\mathrm{H}][\mathrm{PH}_3]\,.
\end{equation}
Assuming $[\mathrm{O}]=[\mathrm{O}]_0$ and $[\mathrm{H}]=[\mathrm{H}]_0$ for all times, and making use of the expression for $[\mathrm{PH}_3]$ in Equation~(\ref{Sol:PH3}), we obtain
\begin{equation}\label{Sol:PH2}
    [\mathrm{PH}_2](t)=\frac{C_{21}}{r_2-r_1}\,e^{-r_1\,t}+C_{22}\,e^{-r_2\,t}\,.
\end{equation}

{\bf Calculation of [PH]:} The differential equation associated with PH is
\begin{equation}
  \frac{1}{n_{\mathrm{H}}}  \frac{d[\mathrm{PH}]}{dt}=-k_{4}[\mathrm{O}][\mathrm{PH}]-k_{7}[\mathrm{N}][\mathrm{PH}]-k_{10}[\mathrm{H}][\mathrm{PH}]+k_{12}[\mathrm{H}][\mathrm{PH}_2]-k_{14}[\mathrm{C}][\mathrm{PH}]\,.
\end{equation}
Assuming $[\mathrm{O}]=[\mathrm{O}]_0$, $[\mathrm{N}]=[\mathrm{N}]_0$, $[\mathrm{H}]=[\mathrm{H}]_0$ and $[\mathrm{C}]=[\mathrm{C}]_0$ for all times, and making use of the expression for $[\mathrm{PH}_2]$ in Equation~(\ref{Sol:PH2}), we obtain
\begin{equation}\label{Sol:PH}
    [\mathrm{PH}](t)=\frac{C_{31}}{r_3-r_1}\,e^{-r_1\,t}+\frac{C_{32}}{r_3-r_1}\,e^{-r_2\,t}+C_{33}\,e^{-r_3\,t}\,.
\end{equation}

{\bf Calculation of [CP]:} The differential equation associated with CP is
\begin{equation}
  \frac{1}{n_{\mathrm{H}}}  \frac{d[\mathrm{CP}]}{dt}=-k_{8}[\mathrm{N}][\mathrm{CP}]-k_{11}[\mathrm{O}][\mathrm{CP}]+k_{14}[\mathrm{C}][\mathrm{PH}]\,.
\end{equation}
Assuming $[\mathrm{N}]=[\mathrm{N}]_0$, $[\mathrm{O}]=[\mathrm{O}]_0$ and  $[\mathrm{C}]=[\mathrm{C}]_0$ for all times, and making use of the expression for $[\mathrm{PH}]$ in Equation~(\ref{Sol:PH}), we obtain 
\begin{equation}\label{Sol:CP}
    [\mathrm{CP}](t)=\frac{C_{41}}{r_4-r_1}\,e^{-r_1\,t}+\frac{C_{42}}{r_4-r_2}\,e^{-r_2\,t}+\frac{C_{43}}{r_4-r_3}\,e^{-r_3\,t}+C_{44}\,e^{-r_4\,t}\,.
\end{equation}

{\bf Calculation of [P]:} The differential equation associated with P is
\begin{equation}
 \frac{1}{n_{\mathrm{H}}}   \frac{d[\mathrm{P}]}{dt}=k_{1}[\mathrm{N}][\mathrm{PO}]-k_{5}[\mathrm{P}][\mathrm{O}_2]-k_{6}[\mathrm{P}][\mathrm{O}\mathrm{H}]-k_{9}[\mathrm{P}][\mathrm{C}\mathrm{N}]+k_{10}[\mathrm{H}][\mathrm{PH}]+k_{11}[\mathrm{O}][\mathrm{CP}]\,.
\end{equation}
We assume $[\mathrm{N}]=[\mathrm{N}]_0$, $[\mathrm{O}_2]=[\mathrm{O}_2]_0$, $[\mathrm{OH}]=[\mathrm{OH}]_0$, $[\mathrm{CN}]=[\mathrm{CN}]_0$, $[\mathrm{H}]=[\mathrm{H}]_0$ and $[\mathrm{O}]=[\mathrm{O}]_0$ for all times. In addition, as mentioned above, we neglect the first term $k_1[\mathrm{N}][\mathrm{PO}]$ because its value is between 10 and 10$^{3}$ times lower than the dominant terms $k_{10}[\mathrm{H}][\mathrm{PH}]$ and $k_{11}[\mathrm{O}][\mathrm{CP}]$. 
From all this, and making use of the expression for $[\mathrm{PH}]$ in Equation~(\ref{Sol:PH}) and the expression for $[\mathrm{CP}]$ in Equation~(\ref{Sol:CP}), we obtain
\begin{equation}\label{Sol:P}
    [\mathrm{P}](t)=\frac{C_{51}}{r_5-r_1}\,e^{-r_1\,t}+\frac{C_{52}}{r_5-r_2}\,e^{-r_2\,t}+\frac{C_{53}}{r_5-r_3}\,e^{-r_3\,t}+\frac{C_{54}}{r_5-r_4}\,e^{-r_4\,t}+C_{55}\,e^{-r_5\,t}\,.
\end{equation}

{\bf Calculation of [PO]:} The differential equation associated with PO is
\begin{equation}
      \frac{1}{n_{\mathrm{H}}}  \frac{d[\mathrm{PO}]}{dt}=-k_{1}[\mathrm{N}][\mathrm{PO}]-k_{2}[\mathrm{N}][\mathrm{PO}]+k_{3}[\mathrm{O}][\mathrm{PH}_2]+k_{4}[\mathrm{O}][\mathrm{PH}]+k_{5}[\mathrm{O}_2][\mathrm{P}]+k_{6}[\mathrm{O}\mathrm{H}][\mathrm{P}]\,.
\end{equation}
Assuming $[\mathrm{N}]=[\mathrm{N}]_0$, $[\mathrm{O}]=[\mathrm{O}]_0$, $[\mathrm{O}_2]=[\mathrm{O}_2]_0$ and $[\mathrm{OH}]=[\mathrm{OH}]_0$ for all times, and making use of the expression for $[\mathrm{PH_2}]$ in Equation~(\ref{Sol:PH2}), the expression for $[\mathrm{PH}]$ and the expression for $[\mathrm{P}]$ in Equation~(\ref{Sol:P}), we obtain
\begin{equation}\label{Sol:PO}
    [\mathrm{PO}](t)=\frac{C_{61}}{r_6-r_1}\,e^{-r_1\,t}+\frac{C_{62}}{r_6-r_2}\,e^{-r_2\,t}+\frac{C_{63}}{r_6-r_3}\,e^{-r_3\,t}+\frac{C_{64}}{r_6-r_4}\,e^{-r_4\,t}+\frac{C_{65}}{r_6-r_5}\,e^{-r_5\,t}+C_{66}e^{-r_6\,t}\,.
\end{equation}

{\bf Calculation of [PN]:} The differential equation associated with PN is
\begin{equation}
      \frac{1}{n_{\mathrm{H}}}  \frac{d[\mathrm{P}\mathrm{N}]}{dt}=k_{2}[\mathrm{N}][\mathrm{PO}]+k_{7}[\mathrm{N}][\mathrm{PH}]+k_{8}[\mathrm{N}][\mathrm{CP}]+k_{9}[\mathrm{P}][\mathrm{CN}]\,.
\end{equation}
Assuming $[\mathrm{N}]=[\mathrm{N}]_0$ and $[\mathrm{CN}]=[\mathrm{CN}]_0$ for all times, and making use of the expression for $[\mathrm{PH}]$ in Equation~(\ref{Sol:PH}), the expression for $[\mathrm{CP}]$ in Equation~(\ref{Sol:CP}), the expression for $[\mathrm{P}]$ in Equation~(\ref{Sol:P}) and the expression for $[\mathrm{PO}]$ in Equation~(\ref{Sol:PO}), we obtain
\begin{equation}\label{Sol:PN}
    [\mathrm{PN}](t)=\frac{C_{71}}{r_7-r_1}\,e^{-r_1\,t}+\frac{C_{72}}{r_7-r_2}\,e^{-r_2\,t}+\frac{C_{73}}{r_7-r_3}\,e^{-r_3\,t}+\frac{C_{74}}{r_7-r_4}\,e^{-r_4\,t}+\frac{C_{75}}{r_7-r_5}\,e^{-r_5\,t}+\frac{C_{76}}{r_7-r_6}\,e^{-r_6\,t}+C_{77}\,e^{-r_7\,t}\,.
\end{equation}

As can be seen above, the solutions for the evolution with time of the abundances of the P-bearing molecules follow a common functional structure (consisting of a sum of exponentials). For this reason, they can be expressed in a more general way as
\begin{equation}\label{abundanciasGeneral}
    [X_i](t)=\left[ \sum^{i-1}_{j=1} \frac{C_{ij}}{r_i-r_j}\,e^{-r_j\,t}\right]+C_{ii}\,e^{-r_i\,t}\,,
\end{equation}
where $[X_i](t)$ are the abundances of species $i$ (index according to Table~\ref{species_numbers}). Constants $C_{ij}$ and $r_i$ depend on the initial conditions of the system (i.e. the initial abundances $[X_i](0)$) and the system parameters (i.e. the reaction rate coefficients $k_{i}$) as follows:  

\begin{eqnarray*}
    r_1  &=&k_{13} n_{\mathrm{H}}  [\mathrm{H}]_0 \\
    r_2  &=&k_{3} n_{\mathrm{H}}[\mathrm{O}]_0 + k_{12} n_{\mathrm{H}} [\mathrm{H}]_0\\
   r_3  &=&k_{4} n_{\mathrm{H}}[\mathrm{O}]_0 + k_{7} n_{\mathrm{H}} [\mathrm{N}]_0 + k_{10} n_{\mathrm{H}} [\mathrm{H}]_0 + k_{14}  n_{\mathrm{H}} [\mathrm{C}]_0 \\
    r_4  &=&k_{8} n_{\mathrm{H}}[\mathrm{N}]_0 + k_{11}  n_{\mathrm{H}}[\mathrm{O}]_0\\
   r_5  &=&k_{5} n_{\mathrm{H}}[\mathrm{O}_2]_0 + k_{6} n_{\mathrm{H}} [\mathrm{O}\mathrm{H}]_0 + k_{9}  n_{\mathrm{H}}[\mathrm{C}\mathrm{N}]_0 \\
    r_6  &=&k_{1} n_{\mathrm{H}}[\mathrm{N}]_0 + k_{2} n_{\mathrm{H}} [\mathrm{N}]_0\\
  r_7  &=&0  \\
    C_{11}  &=&[\mathrm{PH}_3]_0 \\
    C_{21}  &=& k_{13}  n_{\mathrm{H}} [\mathrm{H}]_0 C_{11}\\
    C_{22}  &=&[\mathrm{PH}_2]_0-\frac{C_{21}}{r_2-r_1}\\
    C_{31}  &=&k_{12} n_{\mathrm{H}} [\mathrm{H}]_0\frac{C_{21}}{r_2-r_1}\\
   C_{32}  &=&k_{12} n_{\mathrm{H}} [\mathrm{H}]_0C_{22}\\
     C_{33}  &=&[\mathrm{PH}]_0-\frac{C_{31}}{r_3-r_1}-\frac{C_{32}}{r_3-r_2}\\
    C_{41}  &=&\frac{k_{14} n_{\mathrm{H}} [\mathrm{C}]_0C_{31}}{r_3-r_1}\\
     C_{42}  &=&\frac{k_{14} n_{\mathrm{H}}[\mathrm{C}]_0C_{32}}{r_3-r_2}\\
      C_{43}  &=&k_{14} n_{\mathrm{H}} [\mathrm{C}]_0C_{33}\\
    C_{44}  &=&[\mathrm{PH}]_0-\frac{C_{41}}{r_4-r_1}-\frac{C_{42}}{r_4-r_2}-\frac{C_{43}}{r_4-r_3}\\
    C_{51}  &=&k_{10} n_{\mathrm{H}} [\mathrm{H}]_0\frac{C_{31}}{r_3-r_1}+k_{11}n_{\mathrm{H}}[\mathrm{O}]_0\frac{C_{41}}{r_4-r_1}\\
    C_{52}  &=&k_{10} n_{\mathrm{H}} [\mathrm{H}]_0\frac{C_{32}}{r_3-r_2}+k_{11} n_{\mathrm{H}}[\mathrm{O}]_0\frac{C_{42}}{r_4-r_2}\\
     C_{53}  &=&k_{10}  n_{\mathrm{H}}[\mathrm{H}]_0C_{33}+k_{11} n_{\mathrm{H}}[\mathrm{O}]_0\frac{C_{43}}{r_4-r_3}\\
   C_{54}  &=&k_{11} n_{\mathrm{H}}[\mathrm{O}]_0C_{44}\\
    C_{55}  &=&[\mathrm{P}]_0-\left( \frac{C_{51}}{r_5-r_1}+\frac{C_{52}}{r_5-r_2}+\frac{C_{53}}{r_5-r_3}+\frac{C_{54}}{r_5-r_4}  \right)\\
    C_{61}  &=&k_3 n_{\mathrm{H}} [\mathrm{O}]_0\frac{C_{21}}{r_2-r_1}+k_4 n_{\mathrm{H}} [\mathrm{O}]_0\frac{ C_{31}}{r_3-r_1}+\left( k_5[\mathrm{O}_2]_0  + k_6 [\mathrm{O}\mathrm{H}]_0 \right) n_{\mathrm{H}} \frac{C_{51}}{r_5-r_1}\\
    C_{62}  &=&k_3 n_{\mathrm{H}} [\mathrm{O}]_0 C_{22}+k_4 n_{\mathrm{H}} [\mathrm{O}]_0 \frac{ C_{32}}{r_3-r_2}+\left( k_5[\mathrm{O}_2]_0  + k_6 [\mathrm{O}\mathrm{H}]_0 \right) n_{\mathrm{H}} \frac{C_{52}}{r_5-r_2}\\
   C_{63}  &=&k_4 n_{\mathrm{H}} [\mathrm{O}]_0 C_{33}+\left( k_5[\mathrm{O}_2]_0  + k_6 [\mathrm{O}\mathrm{H}]_0 \right) n_{\mathrm{H}}\frac{C_{53}}{r_5-r_3}\\
    C_{64}  &=&\left( k_5[\mathrm{O}_2]_0  + k_6 [\mathrm{O}\mathrm{H}]_0 \right) n_{\mathrm{H}} \frac{C_{54}}{r_5-r_4}\\
    C_{65}  &=&\left( k_5[\mathrm{O}_2]_0  + k_6 [\mathrm{O}\mathrm{H}]_0 \right) n_{\mathrm{H}} C_{55}\\
    C_{66}  &=&[\mathrm{PO}]_0 -\left( \frac{C_{61}}{r_6-r_1}+\frac{C_{62}}{r_6-r_2}+\frac{C_{63}}{r_6-r_3}+\frac{C_{64}}{r_6-r_4}+\frac{C_{65}}{r_6-r_5} \right)\\
    C_{71}  &=&k_2  n_{\mathrm{H}} [\mathrm{N}]_0\frac{C_{61}}{r_6-r_1}+k_8  n_{\mathrm{H}}[\mathrm{N}]_0 \frac{C_{41}}{r_4-r_1}+k_7  n_{\mathrm{H}}[\mathrm{N}]_0 \frac{C_{31}}{r_3-r_1}+k_9  n_{\mathrm{H}}[\mathrm{CN}]_0 \frac{C_{51}}{r_5-r_1}\\
     C_{72}  &=&k_2  n_{\mathrm{H}}[\mathrm{N}]_0\frac{C_{62}}{r_6-r_2}+k_8  n_{\mathrm{H}}[\mathrm{N}]_0 \frac{C_{42}}{r_4-r_2}+k_7 n_{\mathrm{H}} [\mathrm{N}]_0 \frac{C_{32}}{r_3-r_2}+k_9  n_{\mathrm{H}}[\mathrm{CN}]_0 \frac{C_{52}}{r_5-r_2}\\
    C_{73}  &=&k_2 n_{\mathrm{H}}[\mathrm{N}]_0\frac{C_{63}}{r_6-r_3}+k_8  n_{\mathrm{H}}[\mathrm{N}]_0 \frac{C_{43}}{r_4-r_3}+k_7 n_{\mathrm{H}} [\mathrm{N}]_0 C_{33}+k_9 n_{\mathrm{H}} [\mathrm{CN}]_0 \frac{C_{53}}{r_5-r_3}\\
    C_{74}  &=&k_2 n_{\mathrm{H}}[\mathrm{N}]_0\frac{C_{64}}{r_6-r_4}+k_8 n_{\mathrm{H}} [\mathrm{N}]_0 C_{44}+k_9  n_{\mathrm{H}}[\mathrm{CN}]_0 \frac{C_{54}}{r_5-r_4}\\
    C_{75}  &=&
    k_2 n_{\mathrm{H}} [\mathrm{N}]_0 \frac{C_{65}}{r_6-r_5}+k_9 n_{\mathrm{H}} [\mathrm{C}\mathrm{N}]_0 C_{55}\\
    C_{76}  &=&k_2 n_{\mathrm{H}} [\mathrm{N}]_0 C_{66}\\
    C_{77}  &=&[\mathrm{P}\mathrm{N}]_0-\left( \frac{C_{71}}{r_7-r_1}+\frac{C_{72}}{r_7-r_2}+\frac{C_{73}}{r_7-r_3}+\frac{C_{74}}{r_7-r_4}+\frac{C_{75}}{r_7-r_5}+\frac{C_{76}}{r_7-r_6} \right).
  \end{eqnarray*}
\\

\section{Calculation of the error derived from assuming that the abundant species are constant in the theoretical solution of the system}\label{appendix:approximation}

In our theoretical calculation of the abundance evolution of the P-bearing species (see Appendix~\ref{appendix:analytical}), we assumed that the abundances of those species classified as \textit{abundant} are constant (i.e. $d[X_i]/dt=0$ for C, H, N, O, O$_2$ and OH). This assumption leads to a set of linearized equations and an approximate solution that is almost indistinguishable for all times and temperatures from the solution obtained with numerical methods. In this section we provide a theoretical calculation to explain this striking similarity. In particular, we will show that the error behind this approximation is negligible for the case of one single reaction of the form 
\begin{displaymath}
    \mathrm{A} + \mathrm{B} \rightarrow \mathrm{C} + \mathrm{D}\,,
\end{displaymath} 
where A is a \textit{scarce} species and B is an \textit{abundant} species. 
We do so because the reactants of our set of reactions (see Table~\ref{tab:set_of_reactions} and Figure~\ref{fig:networks}) are always composed of one scarce species and one abundant species (except for reaction 9 which we treated differently). 

For simplicity, the abundances of species A and B will be called $A$ and $B$. According to the law of mass action, their destruction rate is given by the \textit{nonlinear} system $S$
\begin{equation}
    \frac{dA}{dt}=\frac{dB}{dt}=-kAB\,, \quad  A(0)=A_0, \quad B(0)=B_0\,. \label{Eq:ABCD}
\end{equation}
 Our aim is to prove that, when $B_0>>A_0$, the exact solution of the  system $S$ is very similar to the solution of the \textit{linear} system $S'$ in which the abundance of species B is constant, that is: 
\begin{equation}
\frac{dA'}{dt}=-kA'B_0, \quad A'(0)=A_0\,.
\end{equation}
The exact solution of the nonlinear system $S$ is
\begin{eqnarray}
 A(t)&=&\frac{B_0-A_0}{\frac{B_0 e^{(B_0-A_0)kt}}{A_0}-1}\,,\\\label{eq:a2}
 B(t)&=&B_0-A_0+\frac{B_0-A_0}{\frac{B_0 e^{(B_0-A_0)kt}}{A_0}-1}\,,\label{eq:b2}
\end{eqnarray}
and the exact solution of the linear solution $S'$ is 
\begin{eqnarray}
A'(t)&=&A_0 e^{-B_0 kt}\,,\\
B'(t)&=&B_0\,.
\end{eqnarray}
We are interested in the error that we are assuming when we approximate the solution of $S$ by the solution of $S'$ for the abundances of both species A and B. 
Thus we focus on the relative errors of $A(t)$ and $B(t)$ defined as
 \begin{eqnarray}\label{errorA}
   \left| \frac{\Delta A}{A}\right|&=&\left|\frac{A(t)-A'(t)}{A(t)}\right|\,,\\
   \left| \frac{\Delta B}{B}\right|&=&\left|\frac{B(t)-B'(t)}{B(t)}\right|\,.
\end{eqnarray}

To introduce that $B_0 >> A_0$, we define $\varepsilon=\frac{A_0}{B_0}\ll1$\,.
We need to evaluate the relative errors of $A(t)$ and $B(t)$ committed when we approximate the solution of the nonlinear system $S$ by the solution of the linear $S'$ in terms of the parameter $k$ and the initial condition $B_0$ for small $\varepsilon$. To do so, we calculate the first order approximation of their Taylor series and obtain
\begin{eqnarray}
\left|\frac{\Delta A}{A}\right|&=&(-1+e^{-B_0 kt}+B_0 kt) \varepsilon + o(\varepsilon^2)\,,\label{Eq:error1}\\
\left|\frac{\Delta B}{B}\right|&=&(1-e^{-B_0 kt}) \varepsilon + o(\varepsilon^2)\,. \label{Eq:error2}
\end{eqnarray}
We conclude that the solution of the nonlinear system $S$ is well approximated by the solution of the linear system $S'$ (where we assume that the abundances of the species labeled as abundant are constant), as long as the time is not too large and $\varepsilon$ is sufficiently small, that is, when the initial condition of the abundant species $B_0$ is sufficiently larger than the initial condition of the scarce species $A_0$. 

We evaluated these errors for our model's data (which comprise all the reaction rate coefficients $k_i$ and the initial abundances of the reactants compiled in Tables~\ref{tab:set_of_reactions} and \ref{tab:initial_abundances}, the H number density $n_{\mathrm{H}}$ included as $k=k_i n_{\mathrm{H}}$, and $T$=10, 100 and 300 K). We found that the relative errors of A and B verify $|\Delta A/A|<0.003$ and $|\Delta B/B|<0.01$, even for their maximum values in $t=10^5$ yrs. In summary, the calculations developed here for a chemical reaction of the same type as the reactions studied in this work strongly support the main assumption of our theoretical analysis: supposing that the abundances of the abundant species are constant in the total system does not have a significant impact on both abundant and scarce species evolution curves. Finally, note that the second-order or bimolecular reactions that can be treated as first-order reactions (because they can be linearized in the way presented here) are called  \textit{pseudo-first-order reactions} \citep{Chang_2017}) .

  \section{Theoretical analysis of the ratio [PO]/[PN] for low times in the interstellar medium}\label{appendix:ratio}

  The reason of the prevalence of the abundance of PO over PN in many astrophysical environments is still unclear. Our model yields values of [PO]/[PN] $<<1$ for all the scenarios analyzed with the parameters present in Table~\ref{tab:set_of_reactions} and the initial conditions in Table~\ref{tab:initial_abundances} for $t=10^5$ yrs, but for low times PO is systematically more abundant than PN. Here we demonstrate that the abundance of PO at short times is mainly due to the action of reactions 3 (O+PH$_2$ $\rightarrow$ PO+H$_2$) and 4 (O+PH $\rightarrow$ PO+H), while the abundance of PN is mainly a result of reaction 7 (N+PH $\rightarrow$ PN+H). 

  Considering the ratio [PO]/[PN], and applying L'Hôpital's rule in the limit $t \to 0$ because both abundances are zero in the limit, we obtain 
  \begin{equation}
     \lim_{t \to 0} \frac{[PO]}{[PN]}=\lim_{t \to 0}\frac{\frac{d[PO]}{dt}}{\frac{d[PN]}{dt}}\,.\label{Eq:hopital}
  \end{equation}
We substitute Equations~(\ref{eq:total_system_PN}) and (\ref{eq:total_system_end}), which yields
\begin{equation}
     \lim_{t \to 0} \frac{[PO]}{[PN]}= \lim_{t \to 0}\frac{-k_{1} n_{\mathrm{H}}[\mathrm{\mathrm{N}}][\mathrm{PO}]-k_{2} n_{\mathrm{H}}[\mathrm{\mathrm{N}}][\mathrm{PO}]+k_{3} n_{\mathrm{H}}[\mathrm{O}][\mathrm{PH}_2]+k_{4} n_{\mathrm{H}}[\mathrm{O}][\mathrm{PH}]+k_{5} n_{\mathrm{H}}[\mathrm{O}_2][\mathrm{P}]+k_{6} n_{\mathrm{H}}[\mathrm{O}\mathrm{H}][\mathrm{P}]}{k_{2} n_{\mathrm{H}}[\mathrm{\mathrm{N}}][\mathrm{PO}]+k_{7} n_{\mathrm{H}}[\mathrm{\mathrm{N}}][\mathrm{PH}]+k_{8} n_{\mathrm{H}}[\mathrm{\mathrm{N}}][\mathrm{CP}]+k_{9} n_{\mathrm{H}}[\mathrm{P}][\mathrm{CN}]}\,.
       \end{equation}
 Considering the initial abundances, the expression becomes
  \begin{equation}
     \lim_{t \to 0} \frac{[PO]}{[PN]}\approx \frac{k_{3}[\mathrm{O}]_0[\mathrm{PH}_2]_0+k_{4}[\mathrm{O}]_0[\mathrm{PH}]_0+k_{5}[\mathrm{O}_2]_0[\mathrm{P}]_0+k_{6}[\mathrm{O}\mathrm{H}]_0[\mathrm{P}]_0}{k_{7}[\mathrm{\mathrm{N}}]_0[\mathrm{PH}]_0+k_{8}[\mathrm{\mathrm{N}}]_0[\mathrm{CP}]_0+k_{9}[\mathrm{P}]_0[\mathrm{CN}]_0}\,.
  \end{equation}
Here we recall that the abundances of PH and PH$_2$ depend on the parameter $f_{\mathrm{P}}$, since we defined it as the fraction of the total P that is in the form of PH, PH$_2$ and PH$_3$ (while the rest remains in the form of atomic P). In both the numerator and the denominator the terms with PH or PH$_2$ are dominant even for a very small fraction of P in the form of PH and PH$_2$. More precisely, for $f_{\mathrm{P}} \geq 0.01$, these terms are more than 2 orders of magnitude higher than the other terms for all temperatures. Therefore, if we assume $f_{\mathrm{P}} \geq 0.01$ we can neglect the non-dominant terms, obtaining 
   \begin{equation}
     \lim_{t \to 0} \frac{[PO]}{[PN]}\approx \frac{k_{3}[\mathrm{O}]_0[\mathrm{PH}_2]_0+k_{4}[\mathrm{O}]_0[\mathrm{PH}]_0}{k_{7}[\mathrm{\mathrm{N}}]_0[\mathrm{PH}]_0}\,.
  \end{equation}
Although $[\mathrm{PH}]_0$ and $[\mathrm{PH}_2]_0$ depend on the P-hydrogenation fraction $f_{\mathrm{P}}$, note that $[\mathrm{PH}]_0=[\mathrm{PH}_2]_0$ (see Table~\ref{tab:initial_abundances}), and therefore 
   \begin{equation}
     \lim_{t \to 0} \frac{[PO]}{[PN]}\approx \frac{(k_{3}+k_{4})[\mathrm{O}]_0}{k_{7}[\mathrm{\mathrm{N}}]_0}\,.
  \end{equation}

\section{Analysis of the chemical evolution of the system for unequal initial abundances of PH, PH$_2$ and PH$_3$. }\label{appendix:PHs}

Throughout the simulations we have used the same initial abundance for PH, PH$_2$, and PH$_3$ for simplicity (equal to $(f_\mathrm{P}/3) \times 2.57 \times 10^{-9}$, see Table~\ref{tab:initial_abundances}). Here, we study how the chemical evolution of the system changes if these initial quantities are unbalanced. Two different distributions of initial abundances are used: (i) only one PH$_x$ has a non-zero abundance, and (ii) two different PH$_x$ species have the same initial abundance and the other is absent. Note that case (i) is the most biased possible distribution of initial abundances, and therefore the difference obtained between calculations done for this distribution and the one used in the rest of the paper (i.e. the relative error shown in Table~\ref{tab:PHs}) should be an upper bound of any other potential distribution. 

H, PH$_2$, and PH$_3$ are related through the chemical route: PH$_3\rightarrow$ PH$_2\rightarrow$ PH $\rightarrow$ P composed of the very endothermic reaction 13 (H+PH$_3$ $\rightarrow$ PH$_2$+H$_2$), reaction 12 (H+PH$_2$ $\rightarrow$ PH+H$_2$) and reaction 10 (H+PH $\rightarrow$ P+H$_2$). In consequence, as we can see in Table~\ref{tab:PHs}, for low temperatures the system moderately depends on the inequality of initial abundances for PH, PH$_2$, and PH$_3$. On the contrary, for medium and large temperatures the chemical route PH$_3\rightarrow$ PH$_2\rightarrow$ PH $\rightarrow$ P is so efficient that for sufficiently large times the system is in practice independent of the unbalance in the initial abundances of PH, PH$_2$, and PH$_3$. 

\section{Bayesian inference for the reaction rate coefficients with a log-uniform prior distribution}\label{appendix:prior}
\begin{figure*}[t!]
\centering
\includegraphics[width=\textwidth]{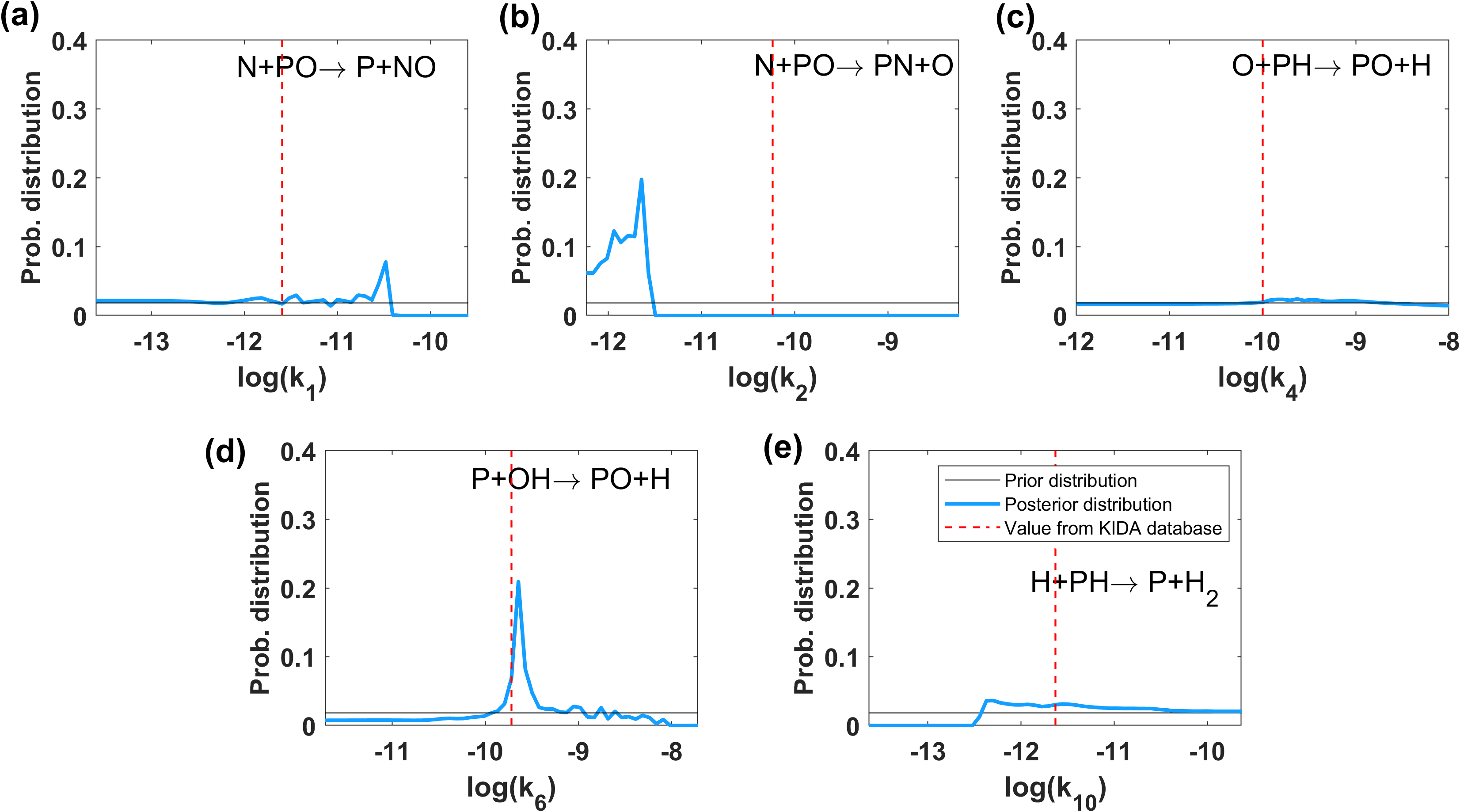}
\caption{Bayesian inference applied to the most important reaction rate coefficients of the model, when log-uniform prior probability distributions are used. (a-e) Prior probability distributions (thin black lines) and posterior probability distributions (PPDs, wide blue lines) obtained with Bayesian inference of the 5 most relevant reaction rate coefficients of our model for $T$=100 K and $t=10^{4}$ yrs, according to observations of star-forming regions from Table~\ref{tab:observations_bayes}. The values of the reaction rate coefficients provided by KIDA and summarized in Table~\ref{tab:set_of_reactions} are plotted (dashed red lines). P-hydrogenation fraction is $f_\mathrm{P}$=0.5 in all calculations.}
\label{bayes2}
\end{figure*}

In Section~\ref{sec:bayes} we used log-normal prior distributions of the reaction rate coefficients $k_i$, according to KIDA guidelines. In order to assess the influence of the prior distribution choice on each $k_i$, we calculate here the posterior probability distributions obtained from a log-uniform prior, that is, a prior devoid of information. The rest of the parameters are as in Section~\ref{sec:bayes} and Figure~\ref{bayes}. Figure~\ref{bayes2} confirms that observational data do not provide relevant information about $k_4$, but shows more clearly than in Figure~\ref{bayes} that $k_1$ has an upper bound ($k_1<3.9\times 10^{-11}$cm$^{3}$ s$^{-1}$) and $k_{10}$ a lower bound ($k_{10}>3.0\times 10^{-13}$cm$^{3}$ s$^{-1}$).

\begin{deluxetable*}{cccccccc}\label{tab:PHs}
\tablecaption{Relative error ($\%$) resulting from the introduction of unequal initial abundances of $\mathrm{PH}$, $\mathrm{PH}_2$ and $\mathrm{PH}_3$ in the calculation of [PO] and [PN] at $t=10^5$ years and $f_\mathrm{P}=0.5$. Two different distributions of initial abundances are used: (i) only one PH$_x$ has a non-zero abundance (equal to $f_\mathrm{P}\times 2.57 \times 10^{-9}$), and (ii) two different PH$_x$ species have the same initial abundance (equal to $(f_\mathrm{P}/2) \times 2.57 \times 10^{-9}$) and the other is absent. }
\tablehead{\colhead{} & \colhead{} & \colhead{$[\mathrm{PH}]_0=0$} & \colhead{$[\mathrm{PH}_2]_0=0$}  & \colhead{$[\mathrm{PH}_3]_0=0$} & \colhead{$[\mathrm{PH}_2]_0=0$} & \colhead{$[\mathrm{PH}]_0=0$} & \colhead{$[\mathrm{PH}]_0=0$} \\
\colhead{} & \colhead{} & \colhead{} & \colhead{}  & \colhead{} & \colhead{$[\mathrm{PH}_3]_0=0$} &\colhead{$[\mathrm{PH}_3]_0=0$}& \colhead{$[\mathrm{PH}_2]_0=0$}
} 
\startdata
 & PO                      & 7.4 \%  & 3.7 \% &   3.7 \% &   15.4 \%&   6.2 \%&   7.3 \% \\
$T$=10 K & PN & 12.1 \% & 20.4 \% & 32.5 \% & 24.4 \% & 41.3 \% & 65.0 \% \\
\hline
 & PO & 0.1 \% & 0.07 \% & 0.05 \% & 2.0 \% & 1.7 \% & 1.7 \% \\
$T$=100 K & PN & 0.1 \% & 0.08 \% & 0.06 \% & 0.6 \% & 1.0 \% & 1.0 \% \\
 \hline
 & PO & 0.02 \% & 0.008 \% & 0.008 \% & 4.9 \% & 4.8 \% & 4.8 \% \\
$T$=300 K & PN & 0.02 \% & 0.008 \% & 0.008 \% & 2.2 \% & 2.2 \% & 2.2 \% \\
\enddata
\end{deluxetable*}

\bibliographystyle{aasjournal}

\end{document}